\documentclass[fleqn,usenatbib]{mnras}

\usepackage[T1]{fontenc}
\usepackage{ae,aecompl}

\usepackage{graphicx}	
\usepackage{amsmath}	
\usepackage{amssymb}	
\usepackage{xcolor}
\usepackage{soul}

\setstcolor{red}
\setul{}{2pt}
\newcommand{\rvir}{r_{200}}
\newcommand{\mvir}{M_{200}}
\newcommand{\vvir}{v_{200}}
\newcommand{\nvir}{N_{200}}
\newcommand{\cvir}{c_{200}}

\newcommand{\zgrp}{z_\mathrm{G}}
\newcommand{\grp}{\mathrm{G}}
\newcommand{\gal}{\mathrm{g}}
\newcommand{\halo}{\mathrm{h}}
\newcommand{\intp}{\mathrm{i}}

\def \g3c {G$^{3}$C\,}
\newcommand{\galform}{\textsc{galform}}
\newcommand{\kids}{KiDS}
\newcommand{\maggie}{\textsc{maggie}}

\newcommand{\LCDM}{$\Lambda{\rm CDM}$}
\newcommand{\zp}{z_{\rm p}}
\newcommand{\zs}{z_{\rm s}}
\newcommand{\zr}{z_{\rm r}}
\newcommand{\galtag}{\textit{galtag}}

\title[Redshift refinement and group richness enhancement]{Galaxy Tagging: photometric redshift refinement and group richness enhancement}

\author[P. R. Kafle et al.]{P. R. Kafle,$^{1}$\thanks{E-mail: prajwal.kafle@uwa.edu.au; pkafauthor@gmail.com}
A. S. G. Robotham,$^{1}$
S. P. Driver,$^{1}$
S. Deeley,$^{2,3}$
P. Norberg,$^{4}$
\newauthor
M. J. Drinkwater,$^{2}$
and L. J. Davies $^{1}$
\\\\
$^{1}$ICRAR, The University of Western Australia, 35 Stirling Highway, Crawley, WA 6009, Australia\\
$^{2}$School of Mathematics and Physics, University of Queensland, Brisbane, Queensland 4072, Australia\\
$^{3}$ARC Centre of Excellence for All-Sky Astrophysics (CAASTRO)\\
$^{4}$Institute for Computational Cosmology, Department of Physics, Durham University, South Road, Durham DH1 3LE, UK\\
}

\begin{document}
\label{firstpage}
\pagerange{\pageref{firstpage}--\pageref{lastpage}}
\maketitle

\begin{abstract}
We present a new scheme, \galtag, for refining the photometric redshift measurements of faint galaxies by probabilistically tagging them to observed galaxy groups constructed from a brighter, magnitude-limited spectroscopy survey. 
First, this method is tested on the DESI light-cone data constructed on the \galform\ galaxy formation model to tests its validity. 
We then apply it to the photometric observations of galaxies in the Kilo-Degree Imaging Survey (\kids) over a 1 deg$^2$ region centred at $15^\mathrm{h}$.
This region contains Galaxy and Mass Assembly (GAMA) deep spectroscopic observations ($i$-band<22) and an accompanying group catalogue to $r$-band<19.8. 
We demonstrate that even with some trade-off in sample size, an order of magnitude improvement on the accuracy of photometric redshifts is achievable when using \galtag.
This approach provides both refined photometric redshift measurements and group richness enhancement. 
In combination these products will hugely improve the scientific potential of both photometric and spectroscopic datasets. 
The \galtag\ software will be made publicly available at \url{https://github.com/pkaf/galtag.git}.
\end{abstract}


\section{Introduction}
Fundamental to many core aspects of galaxy evolution science is the precise and accurate measurement of the distances to galaxies using redshifts. 
There are two largely distinct methods for obtaining these redshifts, either using spectroscopically-identified emission and absorption line features (spectroscopic redshift, $\zs$) or via observed 
broad-band colours matched to a library of spectral templates targeting the large-scale continuum shape (photometric redshift, $\zp$). Due to the nature of spectroscopic observations, 
the former is more precise, but much more observationally costly than the latter. Thus, photometric redshifts can sample orders of magnitude more galaxies for a similar investment of telescope time,
but to a lower fidelity. The trade off between sample size and precision when measuring galaxy redshifts, is largely decided based on the specific scientific question being posed 
(i.e large sample size photometric redshifts for cosmology vs small sample high precision spectroscopic redshifts for group and pair science). However, over the last decade there have 
been vast improvements in the precision of our photometric redshifts based on improved templates, deep and larger are imaging surveys and improvements to photometry fitting algorithms. 
This has led to photometric redshifts becoming big business in the field of galaxy formation and evolution,   \citep[e.g.][etc]{2009ApJ...695..747B,2010ApJ...712..511C,2012amld.book..323B,2013ApJ...775...93D,2017arXiv170609507G},
with survey teams pursuing ever more sophisticated approaches to increase the precision of redshift measurements derived from photometry alone.
 
The different approaches of $\zp$ measurement can be broadly classified into three categories which we discuss below:
\begin{enumerate}
 \item spectral energy distributions (SED)/template fitting technique,
 \item machine learning approach using training and test data, and
 \item inference from cosmic web constraints.
\end{enumerate}

Thus far, the most commonly used technique in $\zp$ estimation is the template fitting methods.
In this method given a library of reference galaxy spectra one fits the observed broadband photometry of a galaxy to find the best fit reference spectra to solve for the redshift.
The completeness of the template and the imperfect observed fluxes due to biases such as disparate zero-point errors in different photometric bands or underestimated errors limits the use of this method. 
An advantage of this method is that it provides fully probabilistic treatment to the redshift measurement, allowing to impose priors over the different types, that can further be a function of redshift, of galaxies.
\cite{1962IAUS...15..390B,1986ApJ...303..154L,1995AJ....110.2655C,1997ApJ...482L..21B,2000ApJ...534..624F,2000ApJ...536..571B,2000A&A...363..476B,2000AJ....120.2206F,2002A&A...386..446L,2008ApJ...686.1503B,2009ApJ...690.1236I,2012MNRAS.421.2355H,2016ApJS..224...24L} etc
are some examples of this category.

In the machine learning approach, first, an empirical model relating galaxy fluxes with redshifts is constructed 
over the training (trustworthy) data for which the exact redshift is already known. 
The trained (predictive) model is then run to predict the redshift of the remaining galaxies (target data).
With the ever increasing efficiency of computers, as well as due to the surge of the spectroscopic spectra from different observational campaigns boosting the sample size of the training data, the machine learning approach has gained more popularity recently. 
An advantage of this method is that during the training phase the model learns the complicated relationships within the observables (e.g. fluxes as a function of redshift which is further a function of galaxy types and so on so forth) that is naturally propagated to the final redshift estimation. 
\cite{2003MNRAS.339.1195F,2009MNRAS.397..520W,2009ApJ...695..747B,2010MNRAS.405..987B,2016PASP..128j4502S,2017ApJ...838....5L,2017MNRAS.466.2039C,2017arXiv170904205B} etc are some examples of this category.

In the third approach, the position of galaxies in physical space and their proximity to large scale structures of the cosmic web are utilised to constrain the redshifts of galaxies.
The applicability of this approach has been limited due to lack of enough overlap between appropriate $\zs$ samples and photometric ones, but where there is overlap it is found to yield good constraint on $\zp$ \citep{2017MNRAS.467.3576M,2017MNRAS.465.1454H}.
This approach is not a stand-alone technique to measure $\zp$, but more of the ancillary approach to calibrate redshift distribution or to further refine the already measured redshifts. 
\cite{2010ApJ...721..456M,2015MNRAS.454..463A,2016MNRAS.460..163R,2017MNRAS.467.3576M} etc are a few examples of this category.
The method we propose in this paper broadly falls into this category, where we will refine the pre-measured $\zp$ using our prior knowledge of the galaxy group distribution.

In this paper we describe a complete implementation of photometric redshift refinement method and present the results of applying the technique to realistic mock catalogues as well as observed data as a proof of concept. 
Throughout the paper, we assume a flat \LCDM\ cosmology with $\Omega_m=0.3$, $\Omega_\Lambda=0.7$, and Hubble parameter $H_0=100\,h\,{\rm km\,s}^{-1},{\rm Mpc}^{-1}$, where we have assumed $h=1$.
This paper is organized as follows. In Section \ref{sec:data}, we describe the GAMA (Galaxy and Mass Assembly) and KiDS (Kilo Degree Survey) observational data as well as the DESI mock catalogue that are used to test our method. In section \ref{sec:method} we outline the halo based prior that is essentially adopted from the \maggie\ \citep[Models and Algorithms for Galaxy Groups, Interlopers and Environment,][]{2015MNRAS.453.3848D}, and the redshift refinement method. In section \ref{sec:results} we show the method in-action. Finally, we discuss and summarize our work and provide future prospects in Section~\ref{sec:summary}.

\section{Data}\label{sec:data}
A minimal data set that is required for our redshift refinement scheme is:
\begin{enumerate}
 \item a galaxy group catalogue constructed on some apparent magnitude limited galaxy redshift survey and 
 \item a galaxy catalogue, fainter than the group catalogue but covering the same area of sky and with photometric redshift measurement which we wish to refine.
\end{enumerate}
To begin with we construct two independent sets of data obtained from disparate sources,
i) a set of observational data includes galaxy catalogue with photometric observations from the KiDS survey $(r>19.8\,\rm{mag})$ and group catalogue from the GAMA survey $(r<19.8\,\rm{mag})$ that share identical sky coverage,
 and ii) a set of theoretical data form the DESI mock catalogue light-cones derived from the \galform\ galaxy formation model. The former forms our test sample to demonstrate the validity of our methods. 
To match the magnitude limit of the observational data, we also divide the DESI catalogue into two parts separated at an apparent magnitude limit on $r = 19.8\,\text{mag}$, identical to that of the GAMA survey.

Below, we provide more details about these data, as well as of the derived quantities.
\subsection{Galaxy and Mass Assembly ({\sc gama}) survey}\label{sec:gama}
\begin{figure}
 \includegraphics[width=1\columnwidth]{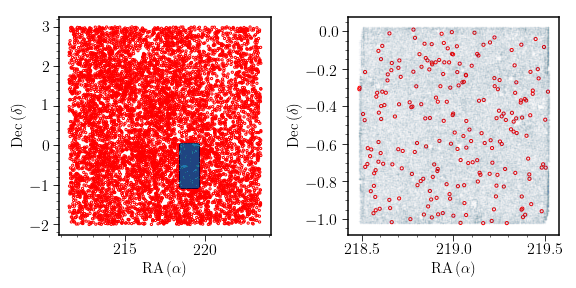}
 \caption{Galaxy and group samples. 
          Left: position of KiDS galaxies in {\sc G15sqrdeg} region (blue region) and overlapping galaxy groups (represented by the positions of the central galaxies) from the GAMA group catalogue (red dots) in the entire G15 region shown in the equatorial coordinates.
          Right: zoomed-in version of the left panel at {\sc G15sqrdeg} region.
          RA and Dec are equatorial angles in degrees.}
 \label{fig:G15deepregion}
\end{figure}
The GAMA survey is a spectroscopic and multi-wavelength survey of $\sim 300,000$ galaxies down to Petrosian $r$-band magnitude $m_r=19.8$ over $\sim286\,\text{deg}^2$ with high spatial completeness carried out on the Anglo-Australian Telescope \citep{2011MNRAS.413..971D,2015MNRAS.452.2087L}. 
Details of the GAMA survey characteristics are given in \cite{2011MNRAS.413..971D}, 
with the survey input catalogue described in  \cite{2010MNRAS.404...86B}, the spectroscopic processing outlined in \cite{2013MNRAS.430.2047H}, and the spectroscopic tiling algorithm explained in \cite{2010PASA...27...76R}, while the group catalogue is provided in \cite{2011MNRAS.416.2640R}.
The group catalogue is constructed using an adaptive Friends-of-Friends (FoF) algorithm, linking galaxies in projected and line-of-sight separations. 
For the full details about the algorithm, diagnostic tests, construction and caveats of the group catalogue we refer the reader to \cite{2011MNRAS.416.2640R}.
As such we only use the galaxy group data from the northern equatorial region of the GAMA survey field centred at $15^{\text{h}}$, i.e., $218.5^\circ<\text{R.\,A.}<219.5^\circ$ and $-1.09^\circ<\text{Dec}<0.0^\circ$ and refer to it as the {\sc G15sqrdeg} region.
In the {\sc G15sqrdeg} region we have 1712 galaxies with $r<19.8\,\rm{mag}$ of which $\sim55\%$ galaxies are present in 236 galaxy groups with richness $\geq2$ whereas remaining galaxies are singleton i.e. with no observed satellites within the magnitude depth of the survey. 
We describe the relevant properties of the group galaxies in Section~\ref{sec:dataprop}. 

The 1 deg$^2$ field centred at G15 region aka {\sc G15sqrdeg} is selected mainly because in this region we have galaxies spectra out to a deeper magnitude limit in $i$-band $m_i=22\,\text{mag}$ than the formal limit of the GAMA survey, providing us with spectroscopic redshifts to compare against our refined photometric redshift and to establish the robustness of our method.
For simplicity, we refer this set of data as a {\sc G15sqrdeg-deep} spectroscopic data.

Spectroscopic observations of the {\sc G15sqrdeg-deep} region were undertaken using the {\sc aat aa\_omega+2df} system in July-Sept 2014. 
Targets were selected to $i<22$ ($r<24$) mag and assigned to fibres using a nightly feedback method, where initially sources were tiled as described in \cite{2010PASA...27...76R}.
Pointings were observed for 40 minute intervals. 
Following each pointing spectra were automatically reduced using {\sc 2dfdr} and assigned redshifts and confidences using {\sc autoz} \citep{2014MNRAS.441.2440B}. 
Sources with secure redshifts were removed from the target list and those without redshifts were re-observed. 
Once multiple observations of the same source were acquired, they were S/N weighted stacked prior to redshifting. 
This process was repeated to allow variable integration times depending on the ability to obtain a redshift for a particular source. 
Once completed, all sources were visually inspected and redshifts adjusted accordingly.    
The catalogue contains 3,241 targeted sources of which 2,289 have a secure redshift ({\sc vis\_class}==`{\sc y}'). 

\subsection{Kilo-Degree Survey (\kids)}\label{sec:kids} 
In the {\sc G15sqrdeg-deep} region we constructed a photometric catalogue of fainter galaxies with $r>19.8\,\text{mag}$ obtained from the Kilo-Degree Survey \citep[\kids,][]{2015MNRAS.454.3500K}.
KiDS is an optical wide-field imaging survey carried out with the VLT Survey Telescope and the OmegaCAM camera. 
To obtain the photometric measurements of {\sc G15sqrdeg-deep} galaxies we undertook following steps. 
First, in the image cut-out centred at the {\sc G15sqrdeg-deep} region we fixed the apertures manually and then measured the photometry using the Lambda Adaptive Multi Band Deblending Algorithm in R (LAMBDAR) software \citep{2016MNRAS.460..765W}. 
LAMBDAR requires at least the image from which one wants photometry measurements and also a corresponding catalogue of aperture parameters. 
Then it places the apertures over the image and measures the flux within them.
Also, it performs deblending for those apertures which intersect with each other and provides the sky background noise to subtract from the galaxies. 
It then estimates noise correlation, calculate flux accounting for local backgrounds. 
Finally, we get fluxes and flux uncertainties over the 4 optical $u, g,r$ and $i$ bands observation from the \kids\ and 5 near-infrared $Z, J, H, K_{\rm s}$ and $Y$ bands from VIKING (VISTA Kilo-Degree Infrared Galaxy Survey).

The complete {\sc G15sqrdeg-deep} photometric catalogue consists of 164,581 galaxies; removing those with incomplete photometric measurements and with $i>22$ mag (to match the magnitude limit of the spectroscopic sample) results in a final sample of 59,134 galaxies. 
The left panel of Fig.~\ref{fig:G15deepregion} shows the entire G15 region of the GAMA survey, where the red dots represent the group central and singleton galaxies whereas the blue mask depict the {\sc G15sqrdeg} region.
The right panel is the zoomed in version of the left panel centred at {\sc G15sqrdeg} region, where blue dots show galaxies in {\sc G15sqrdeg-deep} photometric catalogue.
Next, we use the derived photometry measurements of this sample to estimate their photometric-redshift.

\subsubsection{Photometric redshift measurement}\label{photoz}
In this work we mainly rely on the machine learning approach of ANNz2 \citep{2016PASP..128j4502S} to derive photometric redshifts.
ANNz2 is a new implementation of the code of \cite{2004PASP..116..345C}, which 
utilizes machine learning methods such as artificial neural networks and boosted decision/regression tree, and is freely available software package.
The algorithm uses machine learning methods to learn the relation between photometry and redshift from an appropriate training set of galaxies for which the redshift is already known. 
The trained model is then used to predict the photometric redshift of the galaxies for which spectroscopic measurements are lacking. 

The data we use here to train the ANNz2 networks and generate a catalogue of photometric redshifts consists of galaxies in {\sc G15sqrdeg-deep} region, a subset for which spectroscopic redshifts have been determined (described in Section~\ref{sec:gama}). 
This catalogue consists of 3241 galaxies with $i<22$ mag, out of which 2289 galaxies have a high quality spectroscopic redshift measurement. 
Matching these galaxies up to their corresponding entries in the {\sc G15sqrdeg-deep} photometric catalogue provide us with photometric measurements in the $u,g,r,i,Z,Y,J,H$ and $K_{\rm s}$ bands for most galaxies. 
Removing those with missing or incomplete photometric measurements leaves us with 2,188 galaxies, this being the final sample used in the training and validation runs of ANNz2. 
Half of these galaxies are randomly selected for training with the other half used for validation. 
Finally, we apply the trained ANNz2 networks to the {\sc G15sqrdeg-deep} photometric catalogue to determine their photometric redshifts.

\subsubsection*{Methods}
ANNz2 employs two different approaches which can be selected by the user, namely, Artificial Neural Networks (ANN) or Boosted Decision Trees (BDT). 
Both approaches consist of a training phase where the networks are trained on data with known spectroscopic redshifts, a validation phase and an evaluation phase where the resulting trained networks are applied to a new photometric dataset where the redshifts are unknown. 
In this section we apply both methods and determine which provides the most consistent results for our dataset. 
In both cases we used 50 iterations in the training phase, as additional iterations resulted in limited improvements and increased the risk of biases introduced from over-training, given our limited training sample.

The ANN approach uses at least three layers of nodes, the input layer (consisting of the same number of nodes as the number of input variables), at least 1 hidden layer and a final node which outputs the calculated photometric redshift. 
In each instance of the training run, the number of hidden layers and the number of nodes in each hidden layer are randomly set, along with weightings in the various connections between nodes in neighbouring layers. 
The probability distribution function (pdf) of the galaxy's redshift is determined from the distribution of the weighted photometric redshift estimates from the ensemble of trained networks. 

In contrast, the BDT approach takes the input through an initial root node and passes it through branching linkages of internal nodes before arriving at a final output node, or `leaf'. 
Similarly to the ANN approach, each BDT training run initializes a new tree with different weightings of the input data. 
This results in a `forest' of decision trees, from which the weighted distribution of redshift estimates can be used to determine the pdf for the galaxy's redshift.

\subsubsection*{Training Sample}
First, we look at at the results of the ANNz2 algorithm for the sub-sample with spectroscopic redshifts, and compare the derived photometric redshifts with the spectroscopically determined values. Here we used 50 training runs for both the ANN and BDT methods.

\begin{figure}
\includegraphics[page=1,width=\columnwidth]{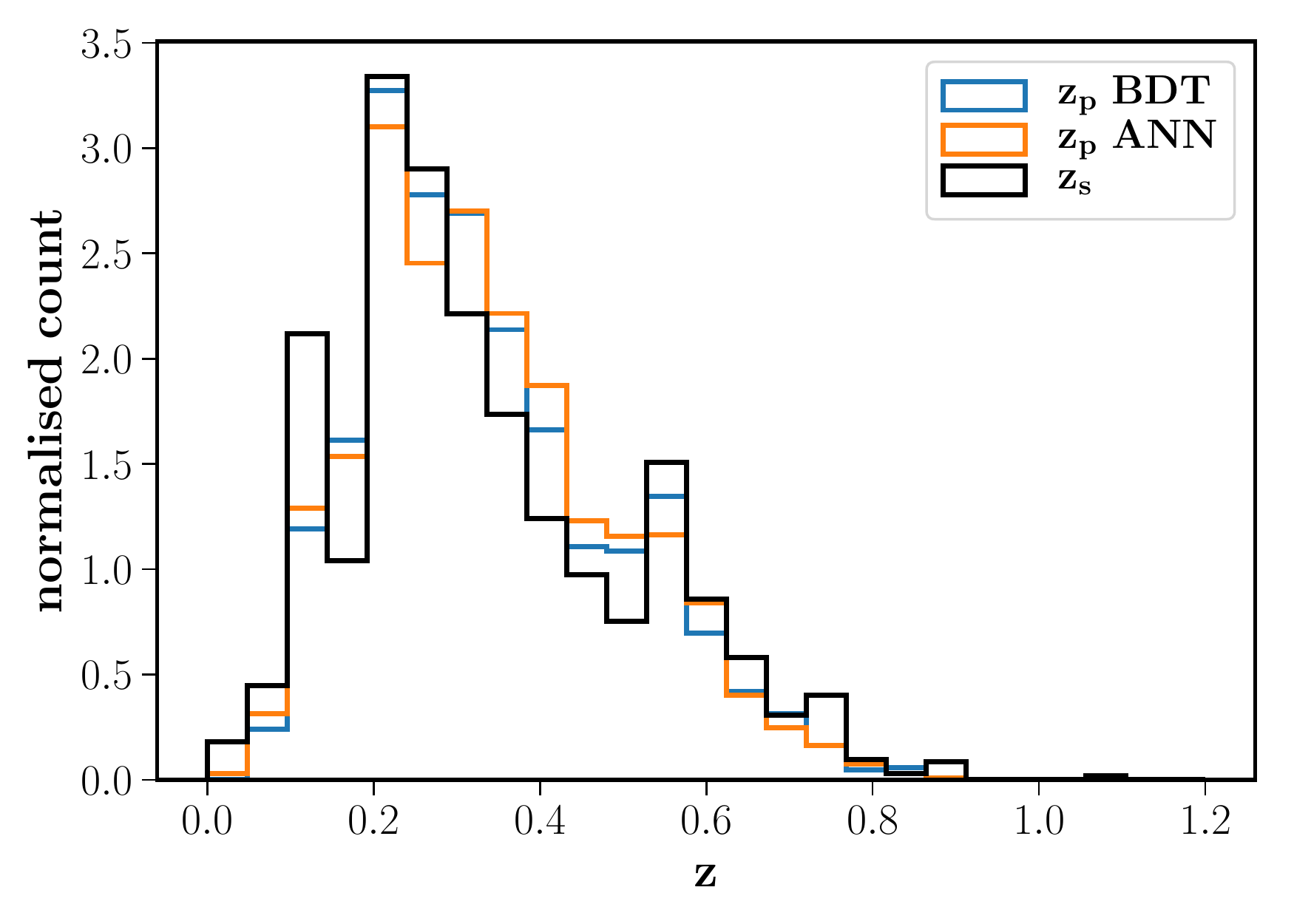}
   \caption{The redshift distributions of the two (ANN and BDT) photometric redshift estimates compared to the spectroscopic redshifts. Both ANNz2 methods produce a similar distribution which follows the spectroscopic distribution.}
 \label{z_distributions}
\end{figure}

\begin{figure}
\includegraphics[page=3,width=\columnwidth]{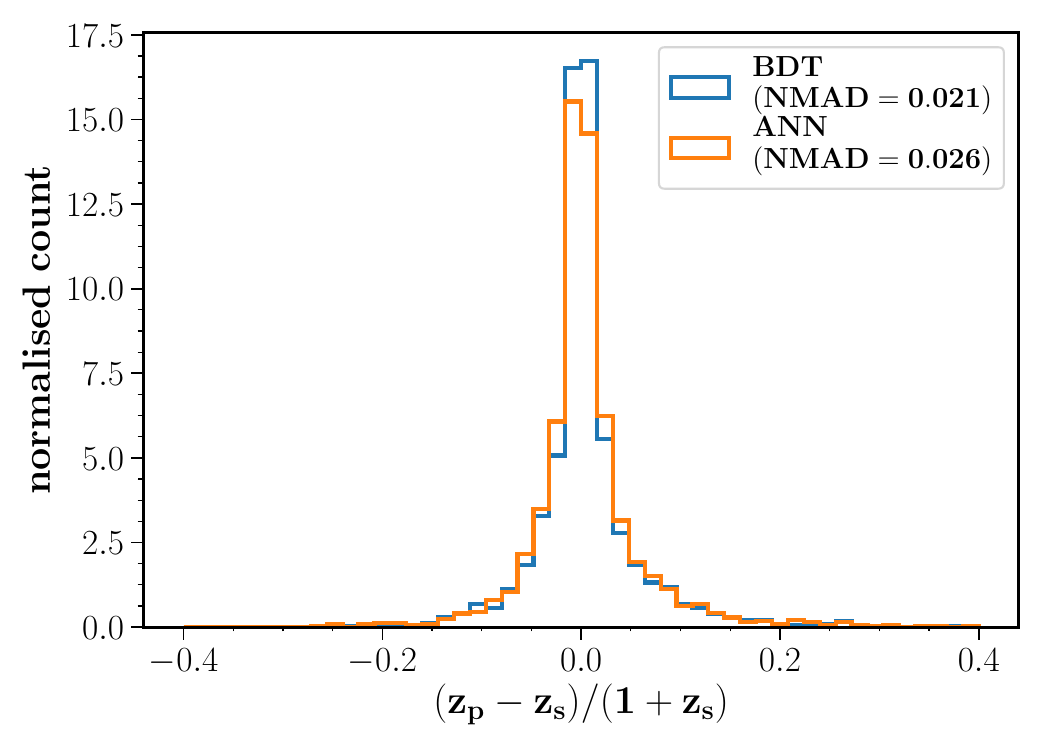}
   \caption{Distributions of scaled biases for the ANN and BDT methods. The BDT distribution, judged from its NMAD value, is marginally better compared to the ANN, indicating more accurate redshift estimates.}
 \label{dz_1plusz}
\end{figure}

\begin{figure}
\includegraphics[page=2,width=\columnwidth]{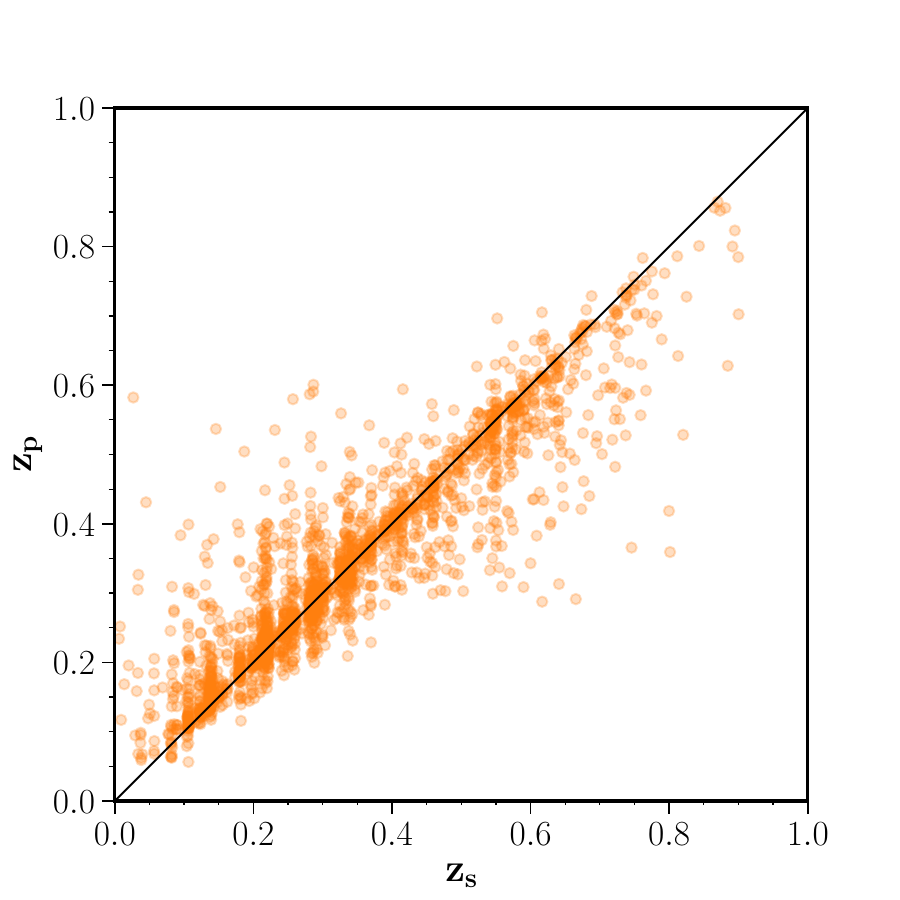}
   \caption{Spectroscopic versus photometric redshifts of the {\sc G15sqrdeg-deep} galaxies produced by the BDT method.}
 \label{photo_spec}
\end{figure}

Fig.~\ref{z_distributions} compares the distributions of the BDT and ANN photometric redshifts $\zp$ with the spectroscopic redshift $\zs$ distribution, highlighting that the overall distribution of redshifts is reproduced well. 
This figure highlights the scattering of galaxies with low $\zs$ values towards higher $\zp$ values, resulting in an under representation of galaxies at low redshift in the photometric distribution. 
Both methods produce $\zp$ which are closely correlated with the spectroscopic value, with the BDT results featuring slightly less scattering. 
However, the distributions for both $\zp$ sets are slightly skewed towards higher values at low redshifts and lower values at high redshifts. 
The scatter is greater at the high end of the $\zs$ due to the small number of training galaxies in this region, and the lower quality photometric measurements for these generally dimmer galaxies. 

The quality of the photometric redshift estimates can be quantified using the normalised median absolute deviation ($\sigma_{\rm NMAD}$ or for simplicity, just NMAD), defined as
\begin{equation}
    \sigma_{\rm NMAD} = 1.48 \times {\rm median} \left( \left\lvert \frac{\Delta z - {\rm median}\left( \Delta z \right)}{1+\zs}\right\lvert \right),
	\label{eq:quadratic}
	\end{equation}
where $\Delta z = \zp - \zs$ and lower NMAD values indicate more accurate redshift estimates. 
The NMAD values for the two ANNz2 methods ANN and BDT are 0.026 and 0.021 respectively. 
These calculations were done using galaxies in the validation set i.e. those not used for training the algorithms. 
Fig.~\ref{dz_1plusz} illustrates the distributions of scaled bias $\Delta z / (1+\zs)$ for the ANN and BDT methods. 
The distribution for the BDT-derived photometric redshifts is more sharply peaked at $\Delta z / (1+\zs)=0$ relative to the other two, furfther indicating that the BDT approach is producing the more accurate redshift estimates. 
For our sample we find that the BDT method gives more accurate redshift estimates than the ANN, with its NMAD statistics comparing favourably to other photometric implementations (see for example \citealt{2013ApJ...775...93D}).  

We also run the template fitting scheme, EAZY \citep{2008ApJ...686.1503B} with empirical templates of \cite{2014ApJS..212...18B} and obtain inferior NMAD value of 0.041 compare to the machine-learning approach. 
We also find that the EAZY photometric redshifts are slightly asymmetric around $\Delta z / (1+\zs)=0$ while both ANN and BDT produce a symmetric distribution. 
Given the $\zp$ distributions with higher NMAD values produced by EAZY and ANN schemes, from this point on we do not consider the results produced by them and only make use of the BDT outputs. 
The relationship between the spectroscopic and photometric redshift for the {\sc G15sqrdeg-deep} spectroscopic catalogue for the BDT approach is shown in Fig.~\ref{photo_spec}. 

\subsubsection*{Probability Distribution Functions}

\begin{figure}
\includegraphics[width=\columnwidth]{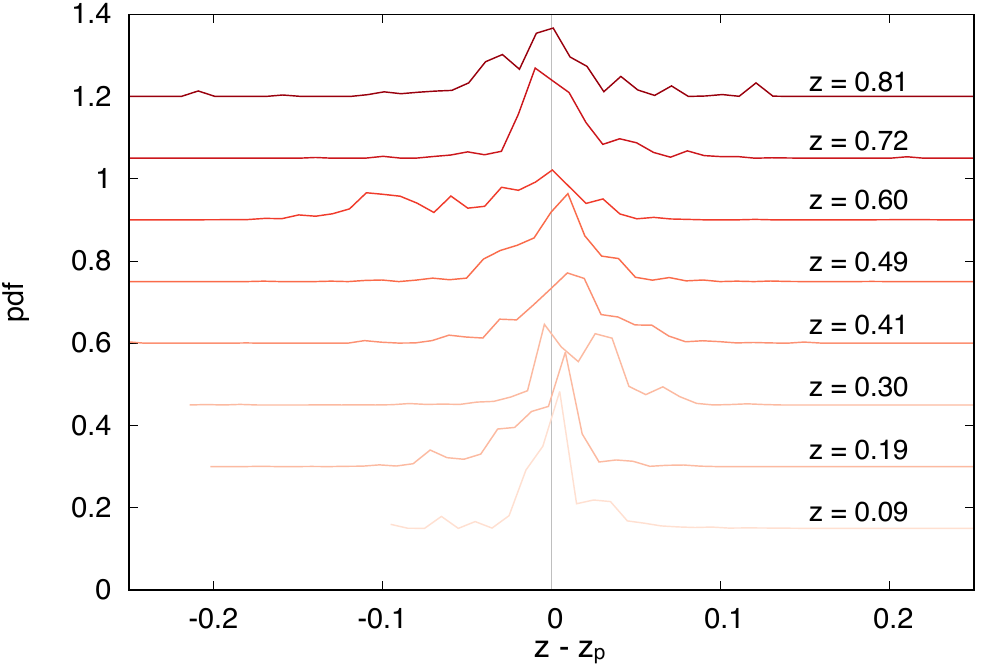}
   \caption{The probability distribution functions found by the BDT method for randomly selected galaxies across the redshift range, centred on the photometric redshift.}
 \label{pdfs}
\end{figure}

In addition to the photometric redshift point-estimate, ANNz2 also produces a probability distribution functions (pdfs) of possible redshifts for each galaxy. 
Fig.~\ref{pdfs} shows a representative sample of the pdfs for galaxy redshifts as determined by the BDT method, with the galaxies taken from the training sample. These pdfs feature strong peaks in most cases, however many pdfs are evidently not very smooth. 
For a minority of galaxies, the pdfs found by the BDT method feature a double peak, though only on rare occasions are the two peaks near equal in amplitude. 
These features may nevertheless have an impact on the next stage of this project, particularly if there is a galaxy group located around the secondary peak.

To get the greatest improvement from the new refinement method, we require pdfs which are unbiased and representative of the actual distribution of true redshifts around the photometric redshifts. The pdfs generated by the BDT approach were tested for uniformity by looking at the $C(z_{s})$ statistic, which gives the total predicted probability that $z_{s}$ was located somewhere between zero and the $z_{s}$ value which was actually observed:
\begin{equation}
    C(z_{\rm s})=\int_{0}^{z_{\rm s}} p(z) dz.
	\label{cz_eq}
	\end{equation}

If the generated pdfs were unbiased and correctly representative of the distribution of $z_{s}$ about $z_{p}$, one would expect to find that 10 percent of galaxies would have measured $z_{s}$ values located in the first 10 percent of their pdf (corresponding to a $C(z_{s})$ value $\leq 0.1$), another 10 percent would have $C(z_{s})$ values between 0.1 and 0.2, and so on. Therefore, finding $C(z_{s})$ for all galaxies in the sample and then finding an empirical cumulative distribution function (ECDF) of all the $C(z_{s})$ values should result in a straight line.

When applying this test to the BDT pdfs, we found that the ECDF deviated from a straight line, indicating that they are indeed biased. Since the deviation was found to be systematic, we corrected for this bias by converting the pdf of each individual galaxy into a cumulative distribution function and, at each point along the distribution, correcting it's value to the corresponding value of the global ECDF. The full details of this correction process are given in a separate paper, Deeley et al. (in prep).

\subsection{Theoretical data}
The DESI light-cone mock catalogues are based on the {\scriptsize GALFORM} galaxy formation model of Gonzalez-Perez et al. (2014). 
The outputs of the model are placed in a light-cone using the technique described in Merson et al. (2013).
The light-cone has a circular field of view of radius 4.0 degrees, and only galaxies with apparent magnitudes brighter than $r \leq 23.8\,\text{mag}$, i.e., 4 magnitudes fainter than the {\sc G15sqrdeg} data, and cosmological red-shifts less than 2.5 are included. 
Similar to the case with the G15 data, here also we construct two separate sub-catalogues, which include 
\begin{itemize}
    \item sets of fainter ($19.8<r/{\rm mag}<19.8+i$) with $i\in 1 \to 4$ galaxy catalogues with synthetic photometric redshift and
    \item a common corresponding halo catalogue with $r\leq19.8\,{\rm mag}$.
\end{itemize}
To estimate the synthetic photometric redshifts for the DESI galaxies, first we take an approach similar to the one for the {\sc G15sqrdeg-deep} data, i.e., given colours and magnitudes employ photometric redshift determination software.
However, we note that irrespective of machine learning and template fitting based photo-z software the yielded photometric redshifts have large variance and significant systematics.
There could be many reasons for this such as imperfect stellar population synthesis models.
To minimise the unknown systematics and have a controlled sample, we generate a pseudo photometric redshift by applying a random error to each mock galaxy redshift randomly generated from a normal distribution. 
This is repeated for normal distributions with two choices of NMAD (or simply, standard deviation as we do not simulate outliers), 0.02 and 0.04, to investigate how the precision of photometric redshift affects out method.   

\subsection{Intricacies of the data: deriving galaxy and group properties}\label{sec:dataprop}
There are a number of key inputs required for our refinement method, including the properties of (i) the group central galaxy (stellar mass and position) (ii) the group (velocity dispersion, virial mass and virial radius), and (iii) the fainter galaxies for which we wish to refine $\zp$ (projected distance from the group centre). Below the derivation of each of these properties is described in detail.

\subsubsection{Group centric distance and velocity}
We consider the brightest galaxy in a group (BGG) as its central galaxy. 
The projected separation ($R$) of a galaxy from the centre of the group are calculated using the cosmological formulae for distance estimation:
\begin{equation}\label{eqn:R}
R = \theta~d_\mathrm{ang}
\end{equation}
where the cosmological angular distance 
\begin{equation}
d_\text{ang}(\zgrp) = \frac{c}{1+z} \int \frac{\mathrm{d} z'}{H(z')},
\end{equation}
$\zgrp$ is the central group galaxy redshift and $c$ is the speed of light.
The angle $\theta$ is the angular separation 
between the galaxy ($\alpha_\gal$, $\delta_\gal$) and central group galaxy 
($\alpha_\grp$, $\delta_\grp$),
where $\alpha$ and $\delta$ are the equatorial 
coordinates representing the Right Ascension and Declination angles respectively. 
Note the projected distance $R$ has to be calculated for all combination of galaxies and central-galaxies. 
Fortunately, $R$ does not depend on the galaxy redshift and therefore, the distance matrix can be calculated once for each data set and later looked-up when needed.  
Similarly, velocity of any galaxy relative to the group centre is given by
\begin{equation}
v/c = \frac{z-\zgrp}{1+\zgrp}.
\end{equation}

\subsubsection{Halo properties}
Finally, we determine the mass of each group using the theoretical relation between central galaxy stellar mass and halo mass, that is one derived from the abundance matching.
A singleton galaxy that is not assigned to any group in a group catalogue could be a potential central galaxy of the group containing unobserved fainter satellites. 
Hence, we also treat a singleton galaxy as a potential group.
From the derived halo mass ($\mvir$) we estimate the group virial radius ($\rvir$) using
\begin{equation}
\rvir = \sqrt[3]{\frac{2 G \mvir}{\Delta H^2(z)}}.
\end{equation}

\begin{equation}
H(z) = H_0 \sqrt{\Omega_m (1+z)^3 + 1 - \Omega_m},
\end{equation}
where $\Omega_m$ is the cosmological density parameter at $z=0$ and the value for the virial over-density parameter $\Delta=200$.
Similarly, virial velocity is calculated using the relation 
\begin{equation}
\vvir = 10 H(z) \rvir,
\end{equation}
whereas concentration parameter $\cvir$ is derived from the concentration-virial-mass relation obtained from 
\cite{2008MNRAS.390L..64D} given by
\begin{equation}
\cvir(\mvir, z_\grp) = 6.71~(0.5\,h\,\mvir/10^{12})^{-0.091} (1+z_\grp)^{-0.44}.
\end{equation}

\section{Method: Galaxy-to-group assignment}\label{sec:method}
We now present the description of the different steps involved in \galtag.
First, we outline the prescription for the phase space distributions of the halo member galaxies and interlopers, where interlopers mean the galaxies that lie outside the virial sphere of the group, but within the cone circumscribing the virial sphere.
Second, we show how galaxies are probabilistically tagged to the potential group.
Finally, we illustrate the photometric redshift refinement process.

We obtain the ansatz for the halo and interloper models from \cite{2015MNRAS.453.3848D,2016MNRAS.458.1301D},
who developed it as a part of the Models and Algorithms for Galaxy Groups, Interlopers and Environment (\maggie).
\maggie\ is a prior- and halo-based abundance matching group finding algorithm, showing a promising alternative to the conventional crispy group-finding scheme such as the friends-of-friends.
For the purpose of our paper we only need and make use of the halo and interloper models given in \maggie\ and not of its group finding aptitude.
While we refer to the above papers for the full derivation, tests and justification of parameters assumed, below we outline minimal complete information that is relevant to our work.

\subsection{Halo surface density}
Following \cite{2013MNRAS.429.3079M}, the density of halo member galaxies $g_\halo (R,v)$ in projected phase-space limited to the virial sphere can be written as
\begin{equation}\label{eqn:ghRv}
g_\halo (R,v) = 2~\int_{R}^{\rvir} \rho(r)~h(v|R,r)~\frac{r}{\sqrt{r^2-R^2}}~\mathrm{d}r,
\end{equation}
where $\rho(r)$ is a galaxy number density profile.
Assuming that a galaxy group is a self-consistent system, i.e. the galaxy number distribution follows the mass distribution we can consider that $\rho(r)$ follows a NFW profile \citep{1996ApJ...462..563N}, given by
\begin{equation}
\rho(r) = \left( \frac{\nvir}{4 \pi \rvir^3} \right) \frac{f(\cvir)}{x(x+1/c)^2} ,
\end{equation}
Here $\nvir$ stands for the number of galaxies within the virial sphere, which as we will see later cancel out and hence, can be assumed to be an arbitrary number at this stage.
Also, $x=r/\rvir$ and the function 
\[f(\cvir) = \frac{1}{\ln(1+\cvir)-\cvir(1+\cvir)}.\]

In equation~\ref{eqn:ghRv}, $h(v|R,r)$ is the probability of observing a line-of-sight velocity at the position (r,R), which is assumed to have a Gaussian distribution written as
\begin{equation}
h(v|R,r) = \frac{1}{\sqrt{2 \pi \sigma_z^2 (R,r)}} \exp \left(-\frac{v^2}{2 \sigma_z^2(R,r)} \right),
\end{equation}
with the squared velocity dispersion run given by
\begin{equation}\label{eqn:sigmaz}
\sigma_z^2(R,r) = \left( 1 - \beta(r) \frac{R^2}{r^2} \sigma_r^2(r) \right).
\end{equation}
Here, 
\[\beta=1 - \frac{\sigma_\theta^2}{\sigma_r^2}\] 
is the velocity anisotropy parameter with $\sigma_r$ and $\sigma_\theta$ 
being the second moments of radial and tangential components of the velocity vector in spherical coordinates relative to the centre of the halo at a rest frame.
It is clear that to calculate $\sigma_z(R,r)$ we must know the $\beta(r)$ and $\sigma_r(r)$ runs of each halo. 
Unfortunately, due to the lack of the peculiar velocity information of galaxies, $\beta$ and $\sigma_r$ are not directly observable quantities.
For this we resort to the theoretical data of the \LCDM\ cosmological simulations,
and choose the following form for the $\beta$ profile 
\begin{equation}
\beta(r) = \frac{r}{2(r + \rvir/\cvir)},
\end{equation}
which is taken from \cite{2005MNRAS.363..705M} and has been shown to agree with a list of different cosmological \LCDM\ simulations. 
Moreover, it is also the recommended $\beta(r)$ profile in \maggie.
When $\beta(r)$ is assumed, we can substitute in the spherical Jeans equation and 
determine the other known unknown, $\sigma_r(r)$.
Thankfully, \cite{2015MNRAS.453.3848D} already provide the solution for us in the set of equations (A1-A5) from the appendix section, which is terms of halo virial properties can be summarised as
\begin{equation}
\begin{split}
\sigma_r^2(r) = & \left(\frac{G \mvir}{\rvir}\right) \frac{\cvir~f(\cvir)}{6y(y+1)} \times \\
& \left[6y^2 (1+y)^2 {\rm Li}_2 (-y) + 6y^4 \coth^{-1} (1+2y) \right.\\
& - 3y^2 (1+2y) \ln y + y^2 (1+y)^2 \{\pi^2 + 3 \ln^2 (1+y)\} \\ 
& \left. -3(-1+2y^2)\ln(1+y)- 3y(1+y)(1+3y) \right],
\end{split}
\end{equation}
where $y = \cvir~r/\rvir$.
Here, ${\rm Li}_2$ is a dialogarithm function defined as
\begin{equation}
  {\rm Li}_2 (-x) =
  \begin{cases}
    \sum_{i=1}^{10} (-1)^i \frac{x^i}{i^2} & x < 0.35\\\\
    -\frac{\pi^2}{12} + \sum_{i=1}^{10} \left( \frac{\ln 2}{i} - \frac{a_i}{b_i}\right) (1-x)^i & 0.35 \leq x < 1.95\\\\
    -\frac{\pi^2}{6} -\frac{1}{2} \ln^2(x) - \sum_{i=1}^{10} (-1)^i \frac{x^{-i}}{i^2} & x \geq 1.95.
  \end{cases}
\end{equation}
\footnote{note, Equation A5 in \cite{2015MNRAS.453.3848D} has a factor of 1/2 missing, and also, the sign shown in the dialogarithm function for $x\geq1.95$ case should be negative}. 
The values for the coefficients $a_i$ and $b_i$ are given in Table A1 of \cite{2015MNRAS.453.3848D}.

\subsection{Interloper surface density}
In their study \cite{2010A&A...520A..30M} analyse the distribution of dark matter particles from a cosmological hydrodynamical simulation and predict that the universal distribution of halo interlopers in projected phase-space can be represented by a Gaussian line-of-sight distribution velocity plus a constant term as follows
\begin{equation}
g_\intp (R,v) = \frac{\nvir}{\rvir^2 \vvir}~\left( A(x) \exp \left[-\frac{1}{2} \frac{(v/\vvir)^2}{\sigma_i^2(x)} \right]  + B \right).
\end{equation}
Calibrating with the galaxies of the semi-analytic model of \cite{2011MNRAS.413..101G} at redshift zero, \cite{2015MNRAS.453.3848D,2016MNRAS.458.1301D} determine that the terms $A$, $\sigma_i$  and $B$ obey the following forms
\begin{eqnarray}
\log(A(x)) = -1.092 -0.01922 x^3 + 0.1829 x^6,\\
\sigma_i (x) = 0.6695 - 0.1004 x^2,~{\rm and}\\
B = 0.0067,
\end{eqnarray} 
where $x = R/\rvir$. 

Finally, utilising the halo member (galaxies within the virial sphere) and interloper (galaxies within the virial cone, but residing outside the periphery of the virial sphere) density distributions, the probability that a galaxy at projected radius $R$ and a relative distance $z$ from the group centre to belong to a given group (to the virial sphere of the real-space group) can be written as
\begin{equation}\label{eqn:grpprior}
  p_\grp(\theta,v|\Theta) =
  \begin{cases}
    \frac{g_\halo(R,v)}{g_h(R,v) + g_\intp(R,v)} & R \leq \rvir\\
    0 & R > \rvir 
  \end{cases}.
\end{equation}
The total assignment probability is non-zero only within the virial cone ($R > \rvir$),
therefore, for a practical purpose the galaxy by central-galaxy dimensional distance matrix (Equation \ref{eqn:R}) has to be only calculated for cases where  $R \leq \rvir$ making it a highly sparse matrix with roughly $95\%$ sparsity.
Here, the distribution is conditioned over $\Theta$, consisting of a set of group properties such as the position of the group centre (RA, Dec, $z_{\text G}$) and group virial properties (primarily, $\mvir$). 
It is to be noted here that the normalization $\nvir$ appears both in the $g_\halo$ and $g_\intp$ distributions, hence, cancels out when we write the probability term $p_\grp(R,v|\Theta)$.

\subsection{Photometric redshift refinement}
From the photometric redshift measurement method we obtain a normalised probability distribution of the galaxy  redshift that can be denoted as $p_\gal(z|z_{\mathsf{t}})$, where $z_{\mathsf{t}}$, a latent variable, is the error free true redshift that can not be observed.
With $p_\gal$ and a model for the galaxy group distribution $p_\grp(\theta,z|\Theta)$ (equation~\ref{eqn:grpprior}), we can express a joint galaxy-group distribution as
\begin{equation}
p(\theta, z|\Theta) = \int p_\grp(\theta,v(z_{\mathsf{t}})|\Theta)~p_\gal(z|z_{\mathsf{t}}) d z_{\mathsf{t}}.
\end{equation}
This allows us to calculate the likelihood for a galaxy to belong to a given group as
\begin{equation}\label{eqn:likelihood}
p_\mathsf{tot} = \int p(\theta,z|\Theta)~{\text d}z,
\end{equation}
which gives a measure of correlation of the galaxy and group redshift distributions.
Finally, the resultant refined probability distribution of galaxy redshift will be given by
\begin{equation}\label{eqn:pref}
p_\mathsf{ref}(\theta,z|\Theta) = p(\theta,z|\Theta)/p_\mathsf{tot}.
\end{equation}
Probabilistically, every galaxy will have some finite probability to belong to all the groups.
But, in the end we aim to find the best-matching galaxy-group pair, that is, to apply a hard assignment. Hard assignment in our case is a two-step process. 
First, we apply a relative criteria, in which we only consider a group for which a galaxy has the highest assignment probability as the best match.
Second an absolute measure, where out of the best-matching galaxy-group pairs we only consider pairs for which the assignment probability is greater than some threshold value.
All the remaining galaxies, with an assignment probability less than a threshold value, are considered un-grouped or a singleton.
The optimal value for the threshold assignment probability is determined from the tests done in the synthetic data as we discuss in the later section.
In cases where we only aim to refine the photometric-redshift, we can skip the second step and for all the existing best-matching pair we can directly calculate the expected value of the redshift for the galaxy given a group using the following formula,
\begin{equation}
\zr = \langle z \rangle=\int_{0}^{z_\mathsf{max}} z~p_\mathsf{ref}(\theta,z|\Theta)~\mathsf{d}z,
\end{equation}
where $z_\mathsf{max}$ can be some arbitrarily large redshift, which should at least accommodate the full range of the $p_\gal(z)$ distribution.
For our fainter galaxies limited to $r<23.8$ mag, $z_\mathsf{max} = 2$ is a large enough value.

\subsection{Group assignment purity}
At this point it is interesting to explore how accurately \galtag\ can assign fainter galaxies. 
Strictly speaking \galtag\ does not assign galaxies to groups, but galaxies have some probability $p_\mathsf{tot}$ (given by equation~\ref{eqn:likelihood}) to belong to the virial sphere of a given group. 
Nevertheless to gauge the accuracy, we hard assign galaxies to the highest probable group and compare the purity of the predicted classification with the true group membership. 
For this we first construct a confusion matrix, which is a square matrix of order two providing the true positive (TP), false positive (FP), true negative (TN) and false negative (FN) counts.
Finally, the group purity fraction or the fraction of correct assignments \footnote{Also known as a rand index} is given by $\frac{\rm{TP+TN}}{\rm{TP+TN+FP+FN}}$, the value for which ranges from 0 to 1, where 1 means all galaxies are correctly assigned to its true group.
This purity fraction can only be calculated for the DESI mock data where we know the true partition for all galaxies. 

\section{\galtag\ in action}\label{sec:results}
The software \galtag\ is written in in Python 2.7 and includes both the halo and interloper models from \maggie\ as well as the refinement step discussed above. The software will be made available at \url{https://github.com/pkaf/galtag.git} \footnote{under GNU general public license (GPL), which guarantees end users the freedom to run, study, share and modify the software.}. Below, we highlight key diagnostic results demonstrating the application of \galtag\ in the DESI mock and {\sc G15sqrdeg-deep} data.

\subsection{With DESI synthetic data}
\begin{figure*}
 \centering
 \includegraphics[page=1,width=2\columnwidth]{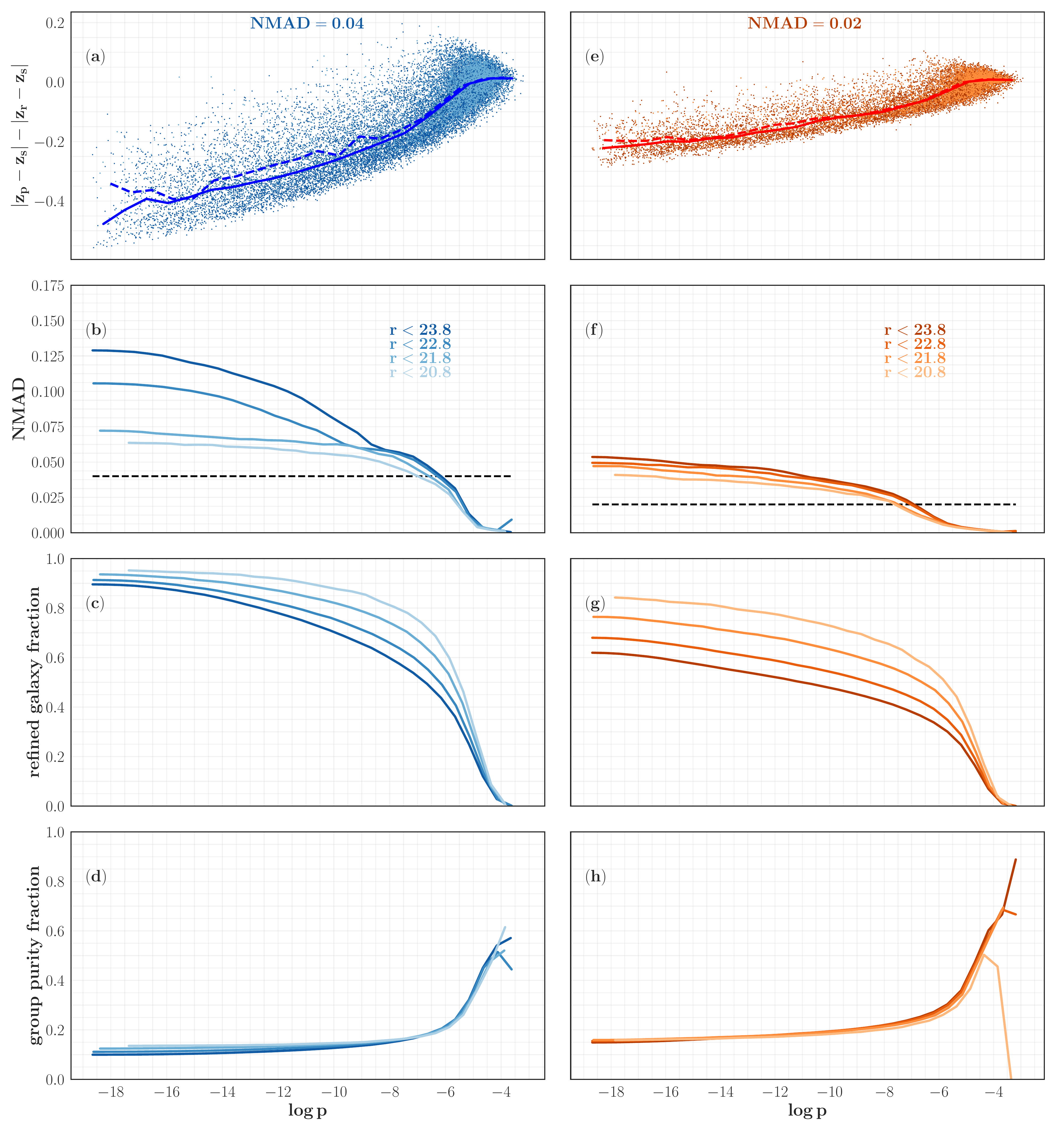}
 \caption{Post-refinement assessment of the DESI data, with intrinsic NMAD=0.04 (left panels) and 0.02 (right panels) as a function of assignment probabilities.}
 \label{fig:desiref}
\end{figure*}

\begin{figure}
\includegraphics[width=\columnwidth]{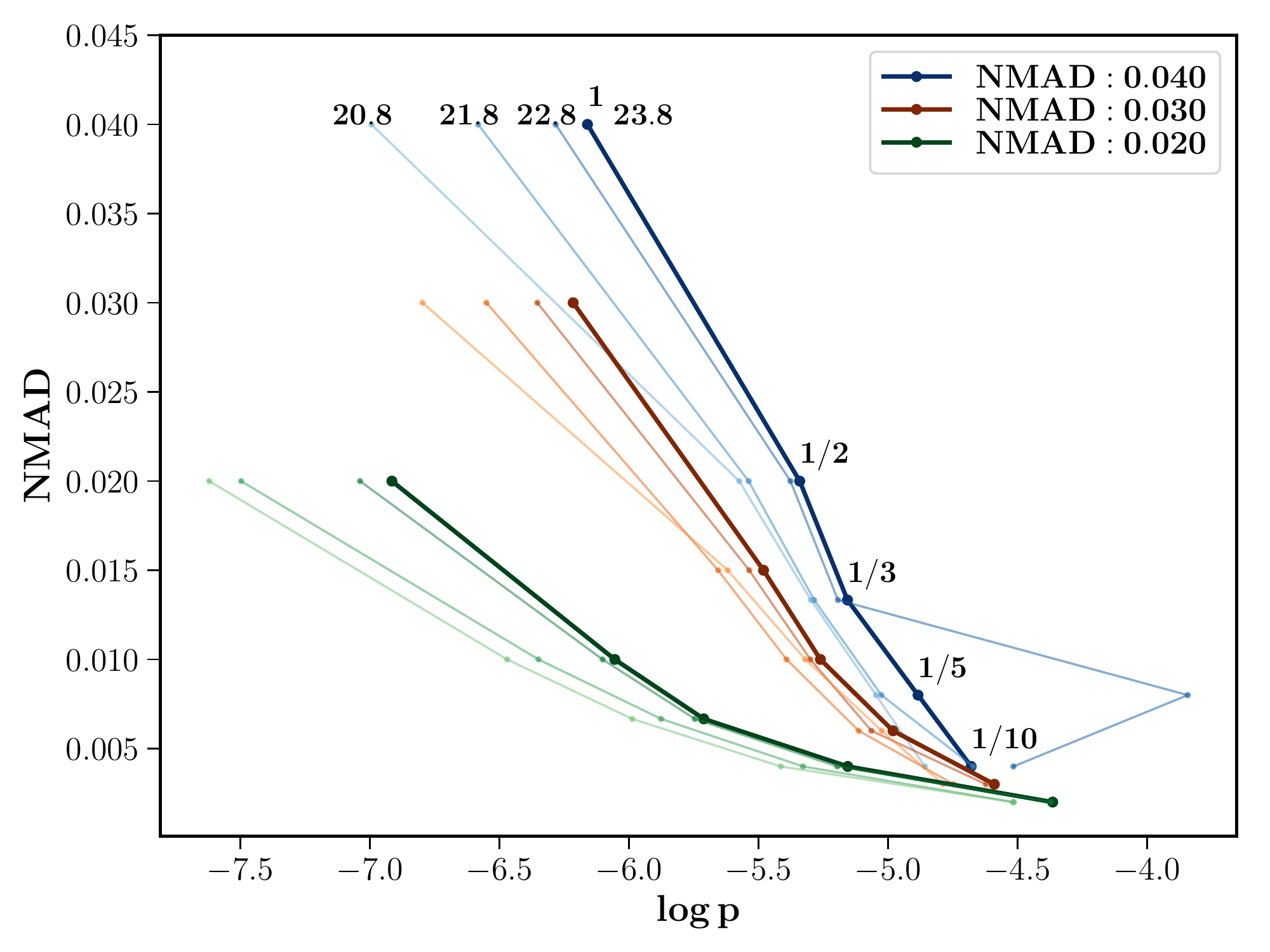}
   \caption{NMAD as a function of minimum probability in the case of DESI data. Different shades of same colour represent data limited to the labelled magnitude limit, where the darkest shades are for cases with $r<23.8\,\rm{mag}$. The labels 1/2, 1/3 etc, representing the fraction of input NMAD, are guides to determine threshold $\log p$ that one should set to obtain corresponding gain in photometric redshift accuracy.}
 \label{fig:nmad_ptot}
\end{figure}

\begin{figure*}
 \includegraphics[page=2,width=2\columnwidth]{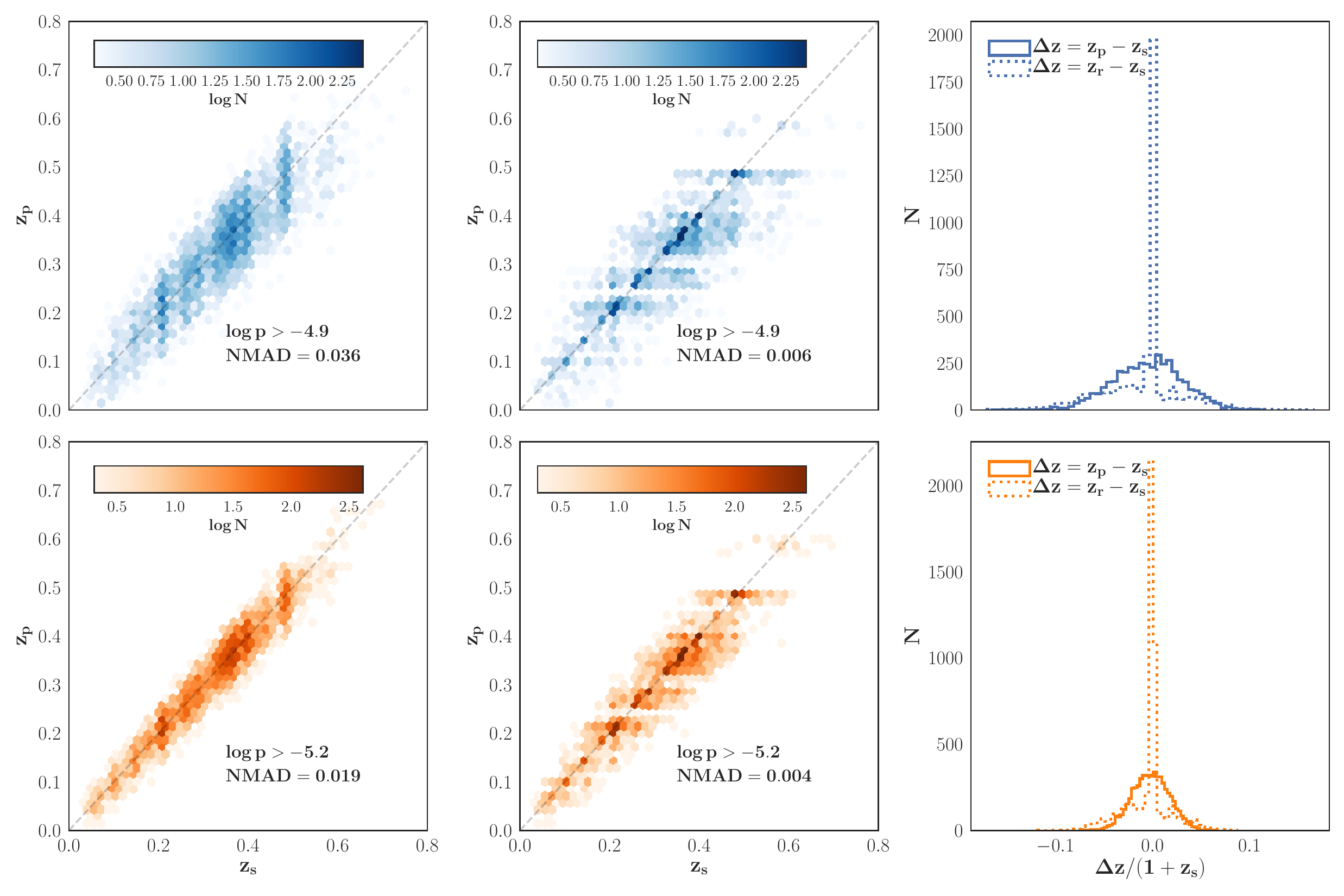}
 \caption{Photometric (right panels) and refined (left panels) redshifts correlations with spectroscopic redshifts, and distributions of scaled-biases before and after refinement (right panels) for the DESI data with intrinsic NMAD=0.04 (top row) and 0.02 (bottom row).}
 \label{fig:desicorr}
\end{figure*}

In Fig.~\ref{fig:desiref} we show the quantitative analysis of refined redshift and group purity for two representative sets of DESI data.
The results in the left panels are for the data with intrinsic NMAD of 0.04 whereas one on the right panels are for the case with NMAD=0.02.
The panels in the top two rows respectively show the biases in redshifts $|\zp-\zs|-|\zr-\zs|$ and the post-refinement NMAD trends as a function of assignment probabilities $(p_\mathsf{tot}=p)$ in the logarithmic scale.
The solid lines of different shades in panels (b) and (f) represent cases with different limiting magnitude ranging between $r<20.8\,\rm{mag}$ to $r<23.8\,\rm{mag}$. 
However, to avoid crowding in the panels (a) and (e) we only show two cases: $r<23.8\,\rm{mag}$ case in darker shade and $r<20.8\,\rm{mag}$ case with fainter shades. 
The solid and dashed lines in these panels represent the respective running medians of $|\zp-\zs|-|\zr-\zs|$ as a function of $\log~p$.
In panels (a) and (e) we observe that only at $\log p\gtrsim-7$ the median biases in redshift measurements are close to zero and at lower probabilities the bias is significantly high.
Moreover, the darker points have much longer low probability tail compare to the fainter points.
Similarly, in panels (b) and (f) we observe that only at $\log p\gtrsim-7$ are the NMAD values of refined redshifts found to improve compared to the intrinsic photometric redshift.
Here, we see that at lower probabilities the NMAD gets much worse than the intrinsic NMAD of 0.04 (left panel) or 0.02 (right panel) and it further worsens with increasing depth of the limiting magnitudes.
These discrepancy are due to the physical effect that most faint galaxies tend to be at larger redshifts, 
and we force them to match groups at low redshift, leading to the underestimation of $\zr$ compare to $\zs$ or $\zp$.   

In panels (c) and (g) we show the fractional cumulative count of the galaxies which have been refined above a given value of $\log p$ whereas in panels (d) and (h) we show the group purity fraction all as a function of $\log~p$.
In both the cases again solid lines with different shades represent cases with different limiting magnitude.
In the figure we see that at $\log p=-7$ we have approximately $50-70$ per cent of the total galaxies that are matched to groups while the group purity fraction is 10 per cent.
However, the trends suggest that for stricter probability cut, with some sample size trade-off, higher group purity is achievable. 

The choice of minimum $\log p$ in \galtag\ is left up to the users so that they can choose based on their science case. In cases where the user only desires the refined redshift, they can set a generous limit. 
However, for projects demanding higher assignment purity one can set a higher threshold probability. 
As a guide in Fig.~\ref{fig:nmad_ptot} we once again present post-refinement NMAD (along y-axis) as a function of $\log~p$ (along x-axis), where we also show an additional intermediate case (NMAD=0.03). 
Different shades of solid lines represent different limiting magnitude whereas different colours display different intrinsic NMAD.
The dots from left to right in each case can be used to infer the threshold $\log~p$ for which NMAD can be improved by 1, 1/2, 1/3, 1/5 and 1/10 times the intrinsic NMAD.
We see that, for example, in a case with intrinsic NMAD=0.02 and the limiting magnitude of $23.8~\rm{mag}$
setting $\log p\simeq-4.5$ will give an order of magnitude improvement in the NMAD value. 
For this case the group purity fraction is approximately 60 per cent and is comparable to the $\sim80$ per cent halo assignment accuracy from the input group catalogue \citep{2011MNRAS.416.2640R}.
The fraction of refined sample compare to the total number of galaxies within the limiting magnitude in this case is only $10$ per cent, which on its face value seem small. 
However, this forms the 85 per cent of the total sample of galaxies within the group redshift range and these are the only galaxies for which we expect any improvement. 

In summary, to highlight the improvement in the redshifts from our refinement, Fig.~\ref{fig:desicorr} we show the quintessential redshift correlations between $\zp-\zs$ (left panels) and $\zr-\zs$ (mid panels).
The top row shows the case of input NMAD=0.04 whereas the bottom row shows the case of input NMAD=0.02.
Here, we have only shown the galaxies with assignment probability $\log~p>-4.9$ ($-5.2$),
the probability at which the NMAD value post-refinement is $1/5^{\rm{th}}$ compared to the intrinsic value. 
The darker points in the mid-panel, along the 1:1 correlation line that represent higher density of points, is enhanced compare to the left panel.
This qualitatively shows the improvement in redshift values due to refinement. 
The right most panels show the distributions of the scaled bias where solid lines are for biases in photometric redshifts whereas dashed lines show the biases in the refined redshift, both compared to the spectroscopic measurements.
As expected, we see that the distributions for scaled refined redshifts are much narrower and peaky in comparison with the scaled photometric redshift distributions.
Note, the wings of the distributions of the scaled biases in the refined cases are slightly asymmetrical due to the underestimation of $\zr$, as discussed earlier in Fig.~\ref{fig:desiref} (a) and (b), which can be eliminated by imposing stricter $\log~p$ cut.

\begin{figure}
\centering
 \includegraphics[page=3,width=1\columnwidth]{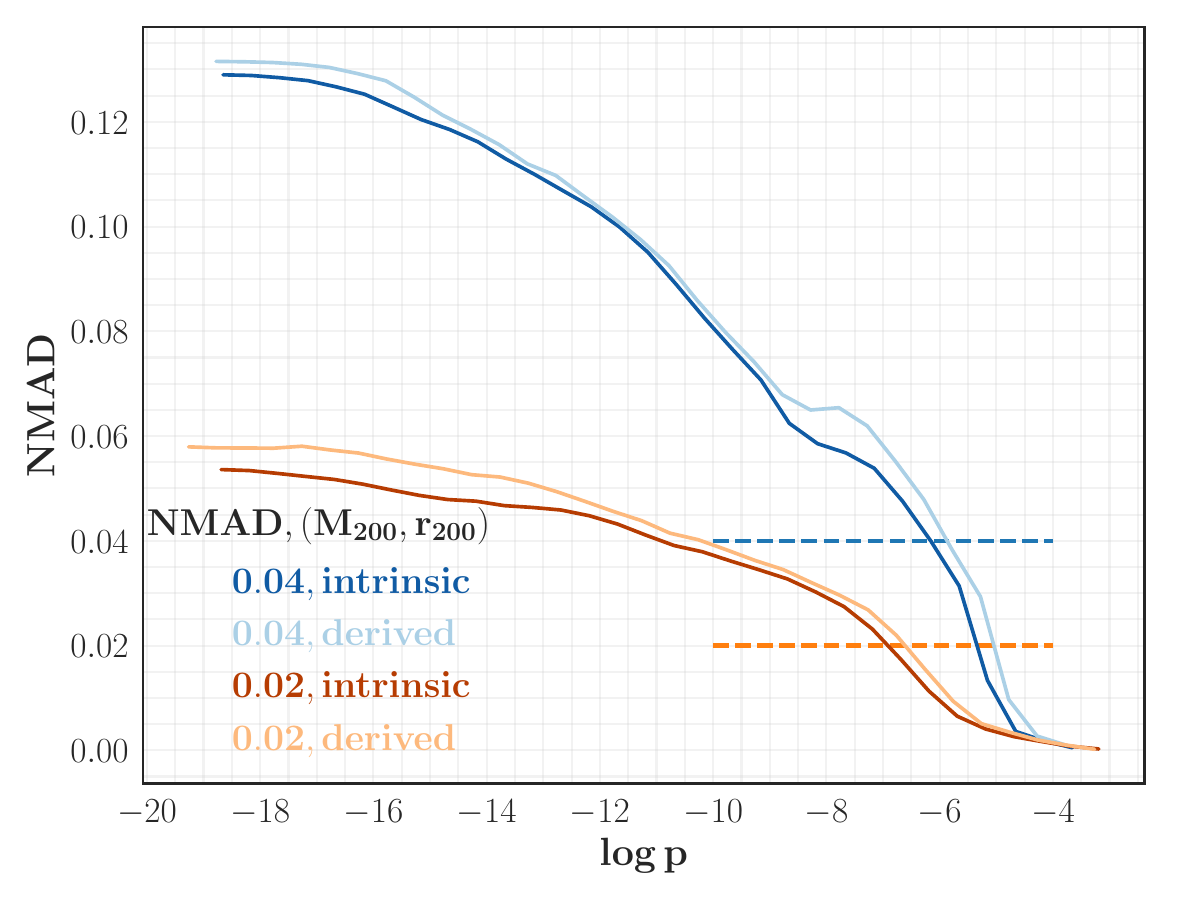}
 \caption{NMAD of the refined redshift of the DESI data generated with intrinsic NMAD=0.04 (blue lines) and 0.02 (orange lines) as a function of assignment probability 
          for the two cases of intrinsic (solid lines) and derived halo virial properties.}
 \label{fig:desihalodef}
\end{figure}
We re-run the analysis for the DESI mock data with a different definition of the halo virial masses  to understand its effect on our final result. 
The results from this additional exercise is shown in Fig.~\ref{fig:desihalodef}.
The figure shows the relation between derived NMAD from the refined redshift sample as a function of $\log~p$ for two different definitions of halo virial properties.
Here again the blue and orange lines represent the sets of DESI data with photometric NMAD values of 0.04 and 0.02 respectively.
The darker lines are when we consider the intrinsic halo virial masses and radii provided by the halo catalogues whereas the fainter lines are when we use the values of halo virial properties derived from the line-of-sight velocity dispersion of group members. 
We observe that at any assignment probability NMAD values for the intrinsic case is always slightly smaller than for the derived case suggesting that the results obtained from the former case is marginally better.
Importantly, the improvement is marginal, which allows us to confidently apply our method to the real data where virial properties are largely inferred from the group velocity dispersions.

\subsection{With {\sc G15sqrdeg-deep} data}
\begin{figure}
\centering
 \includegraphics[page=1,width=1\columnwidth]{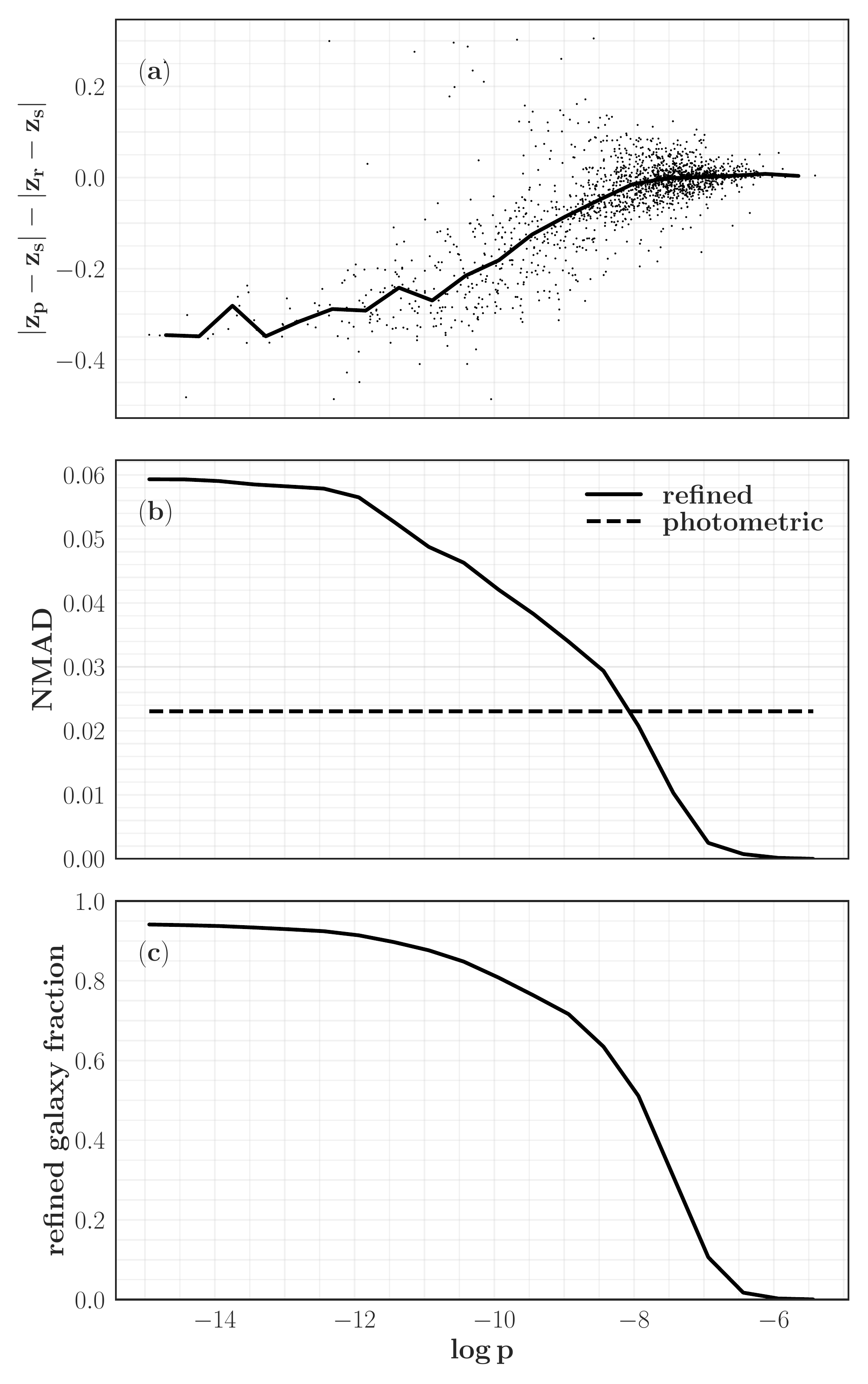}
 \caption{Post-refinement assessment of {\sc G15sqrdeg-deep} data as a function of assignment probabilities. Labels are similar to Fig.~\ref{fig:desiref}.}
 \label{fig:g15ref}
\end{figure}

\begin{figure*}
\centering
 \includegraphics[page=2,width=2\columnwidth]{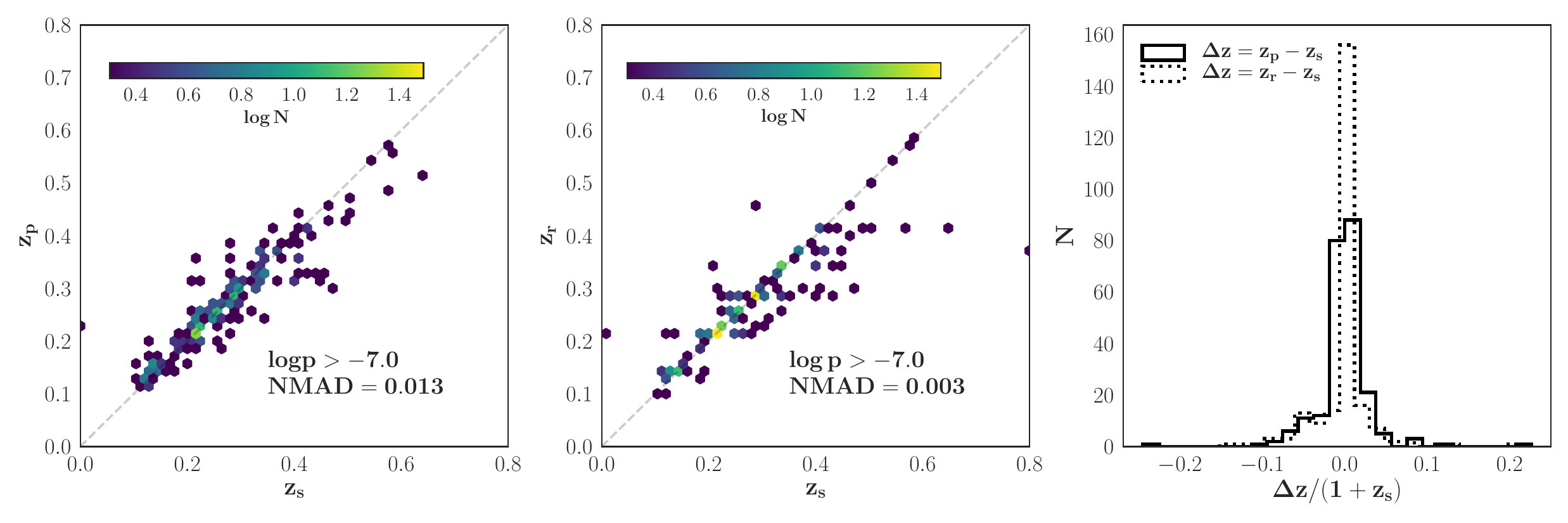}
 \caption{{\sc G15sqrdeg-deep} spectroscopic versus photometric (left panel) and refined (centre panel) redshifts, and distributions of scaled-biases before and after refinement (right panel).}
 \label{fig:g15corr}
\end{figure*}

Similar to the DESI mock data, we also process the {\sc G15sqrdeg-deep} data with \galtag, and present the results in Fig.~\ref{fig:g15ref}. 
We observe trends consistent to one observed in the DESI mock data. 
Such as the median bias $|\zp-\zs|-|\zr-\zs|$ shown with black solid line in panel (a) ceases to zero at larger values of $\log~p$.
Also, as shown in panel (b) the NMAD value for the refined data (shown in black solid line) improves at larger values of $\log~p$. 
For sufficiently large cut-off values for $\log~p$, we can see that even an order of magnitude gain in NMAD values is achievable.
Additionally, the panel (c) shows the fraction of refined galaxy again as a function of $\log~p$. 
Furthermore, to give the sense of improvement in the redshift measurements, in Fig.~\ref{fig:g15corr} we show the redshift correlation between the $z_p$, $z_s$ and $z_r$.
Here we have only considered galaxies that have group matching probability of $\log p > -7$,
resulting reduction of NMAD by $1/5^{\rm{th}}$. 
The improvement in redshift measurements post-refinement can also be gauged from the enhanced number density at 1:1 correlation line seen in the mid-panel compare to the left-most one. 
Similarly, the leaner and peaky distribution of scaled bias $(\zr-\zs)/(1+\zs)$ compare to the distribution of the $(\zp-\zs)/(1+\zs)$, shown in the right-most panel, also demonstrate the improvement achieved post-refinement. 
We note that the cross-over point, that is point where NMAD value for refined sample is same as the value for photometric sample, happens at $\log p \simeq -8$, which is achieved sooner than in the case of the DESI data. 
This is akin to observational uncertainties in various derived quantities that the observed data possess, which get propagated to the final measurements of $\log p$ values.

\section{Discussion, summary and science exploitation}\label{sec:summary}
Before we summarize, we would like to point out the main limitations of our work.
The input photometric redshift and group catalogues both have their own caveats that \galtag\ will naturally inherit. 
For example, \cite{2011MNRAS.416.2640R}, in their studies of GAMA mock catalogues conclude that the halo assignment accuracy with spectroscopic redshifts is only $\sim80$ per cent.
Furthermore, the accuracy of low occupancy groups ($<5$ members) worsens to $\lesssim50$  per cent, and moreover, they form the significant $\sim90$  per cent of the total group population. 
Fidelity of input group catalogue is just the first tier of the issue which is further complicated by the need to define and estimate contentious quantities such as the centre and global group properties.
There is no unique way to define the group centre, as such any of either centre of mass/light, geometric centre, or brightest group galaxy can be considered as a group centre. 
`Correct' selection of a group centre is crucial particularly for very low-occupancy groups (say with $\lesssim3$ members) where perhaps the only other general way to estimate the halo mass is to map their central galaxy stellar-mass to halo-mass assuming the theoretical stellar-mass--halo-mass relation. 
Approximately $60\%$ of galaxies in GAMA within a magnitude limit of $r<19.8~\rm{mag}$ are singletons and are potential group centre for fainter satellite galaxies.
Again, we have to use the theoretical stellar-mass--halo-mass relation to predict halo masses for singletons. 
In order to predict the concentration parameter from the group virial mass, yet another theoretical relation we have to assume is the concentration-virial mass relation.
The above discussed  theoretical  predictions  are  for  pure  dark  matter simulations,  and  are  prone  to  serious  systematics  as they do not include baryonic processes such as cooling, star formation, and feedback. 
For example, the collapse of gas due to  cooling  leads  to  adiabatic  contraction  of  the  dark  matter  halo,  which  increases  its  concentration.  
Feedback,  on the  other  hand,  can  have  the  reverse  effect.
Also, it has been observed that even in the cases of the Milky Way and M31, galaxies that can be studied in greatest details, the derived concentration-virial mass relations do not agree with the theoretical prediction \citep{2014ApJ...794...59K,2018MNRAS.475.4043K}.
Similarly,  the relationship  between  dark  matter  halos  and  galaxy  stellar masses from the halo abundance matching technique rely on the accuracies of observed stellar mass function, the theoretical halo mass function and techniques of abundance matching.
However, for groups with high number of occupants, the line-of-sight group velocity dispersion can provide unbiased and robust handle on the dynamical mass of the groups even in the case of weak perturbations in group membership \citep{1990AJ....100...32B}.
That being said, \cite{2011MNRAS.416.2640R} find that in 80 per cent of all mock groups the recovered velocity dispersion is only within $\sim50$ per cent of the true value and are as likely to have underestimate as overestimate of the velocity dispersion.
Furthermore, our scheme is not provisioned to discover any new groups whose even central galaxies were unobserved in the input magnitude-limited group catalogue. 
Both the central and satellite members of such groups are either matched to observed groups or left unassigned, determined by probability cuts. 
The $\zp$ of galaxy members of such groups can still get improved, provided they are correctly matched to group that is nearby in physical space. 
However, inability to identify such fainter group as a stand-alone individual group will hamper the completeness of group catalogue resulting from the matching process. 
However, this should not have significant impact on group-by-group studies, provided the fainter photometric galaxies are constructed out of surveys with high spatial completeness.
Therefore, we can summarise with the remark that even at best the accuracy of processed redshifts and groups matching are limited by the pitfalls of input catalogues, like in the case of any other scientific exploitation of a group and/or photometric redshift catalogues. 

In addition to the observational limitations discussed above, our approach is also likely to be impacted from the modelling approximations that we have to make.
As we have highlighted earlier, the probabilistic halo model that we adopt for group matching are of \cite{2015MNRAS.453.3848D}, who conduct extensive tests with the cosmological simulations to establish the model.
However, the veracity of a few fundamental assumptions made in the model can still be questioned.
For example, the three-dimensional velocity distribution of the galaxies in \LCDM\ haloes
are not strictly Gaussian \citep{2005MNRAS.361L...1W}. 
Similarly, the model assumes the galaxy number density distribution within the haloes to have the NFW \citep{1996ApJ...462..563N} form, while \cite{2004MNRAS.349.1039N} suggest the Einasto model to be a better fit. 
Moreover, the velocity anisotropy profile is assumed to represent the particles in cluster mass \LCDM\ haloes \citep{2005MNRAS.363..705M}.
Firstly, the galaxy groups are considered to be a less evolved object and whether they follow the similar dynamics to that of the rich clusters and whether our knowledge about the clusters can be scaled and extended to groups or not is still an open question.
Secondly, in any case, the velocity anisotropy is not directly observable for the galaxy groups as we can only measure the line-of-sight component of the velocity vector of the galaxies in groups.
Therefore, the correctness of the assumed profile is unknown.
An incorrect assumption about the velocity anisotropy could lead to the notorious mass-anisotropy degeneracy, meaning underestimation of the anisotropy results the overestimation of groups mass profile and vice-versa.

This is a proof of concept paper that present a new scheme \galtag\ to refine galaxy photometric redshift and enhance group membership based on our prior knowledge of the galaxy group distribution. Here, we forgo an explicit conclusion as we attempt to summarise the paper in the abstract. 
However, we like to briefly discuss potential scientific objectives of the project that we will pursue in future works. 

In a forthcoming paper (Kafle et al. 2018, in prep) we aim to extend the method to two independent sets of observed data namely, the $\sim300$ deg$^2$ of KiDS data overlapping the entire GAMA fields and the $\sim6\,{\rm deg}^2$ of DEVILS (The Deep Extragalactic Visible Legacy Survey \url{https://devilsurvey.org/wp/}) data overlapping the COSMOS fields \citep{2007ApJS..172...99C}. 
The key science we will carry out with this new extended group catalogue is a robust measurement of the galaxy occupation frequency for a large dynamic range of stellar mass and halo masses, including Local Group mass systems down to sub 1/10 times Magellanic Cloud mass galaxies.
Having an $r<19.8~\rm{mag}$ group catalogue over $\sim300$ deg$^2$ and populated additional satellites by
adding photo-z galaxies with the \galtag\ method described above, we will use this combined data to probe significantly further down the luminosity distribution for a large range of halo masses. As well as measuring the luminosity distribution for fainter satellites, the data will also allow for a much more accurate measurement of the luminosity distribution throughout the full range of halo masses that host galaxies. This will place the Milky-Way halo in context, and provide new data for modern galaxy formation models.
Furthermore, by being able to measure the luminosity distribution of individual halos rather than a statistical stack (which is the approach used in clustering based halo occupation distribution work) we will be able to identify whether any individual groups share the luminosity distribution characteristics of the Milky-Way halo. This is important since we might well find that the distribution of the faintest satellites is more or less likely given the presence of the bright Magellanic satellites. As well as pushing the direct measurement of the halo luminosity distribution into a new regime, this dataset will also open up numerous fresh avenues of scientific exploration. Future work could explore the stars, dust, gas, shape, colour, structure and spatial distribution of low mass satellites. All of this information is available to GAMA and already exists for Local Group dwarf
galaxies, opening up multiple future avenues of comparative exploration.
In short, we can utilise the data products for projects that do not require exemplary redshift such 
as to address the missing satellite problem \citep{1999ApJ...522...82K}, the too-big-to-fail problem \citep{2011MNRAS.415L..40B,2012MNRAS.422.1203B} and also to test the lopsidedness of satellite galaxy systems \citep{2016ApJ...830..121L,2017arXiv171007639P} as well as to look for the Local Group analogues to put our Milky Way and neighbouring galaxies in a cosmological context \citep{2012MNRAS.424.1448R,2017arXiv170506743G}.

Beyond comparisons to the Local Group, this new assortment of halo luminosity distributions will
serve as a key reference point for future simulation and theory work. By combining the data in the
manner described we can do much more than present a simple `average' luminosity distribution, in-
stead we will also measure the allowed distribution space that individual halo luminosity distributions
are allowed to occupy. This will allow us to characterise sub-populations for different halo masses,
information that is entirely lost with current statistical stacking techniques and in broad-brush halo occupation distribution techniques.

\section*{Acknowledgements}
PRK is funded through Australian Research Council (ARC) grant DP140100395 and The University of Western Australia Research Collaboration Award PG12104401 and PG12105203. We like to thank Violeta Gonzalez-perez for providing us DESI light-cones, Gary Mamon (IAP) for \maggie\ related Q\&A, Maciej Bilicki for comments on the photometric redshift aspects, and Rob Finnegan and Dylan Cusack-Paquelet for their supports during the earlier stage of the project.

GAMA is a joint European-Australasian project based around a spectroscopic campaign using the Anglo-Australian Telescope. 
The GAMA input catalogue is based on data taken from the Sloan Digital Sky Survey and the UKIRT Infrared Deep Sky Survey. 
Complementary imaging of the GAMA regions is being obtained by a number of independent survey programmes including GALEX MIS, VST KiDS, VISTA VIKING, WISE, Herschel-ATLAS, GMRT and ASKAP providing UV to radio coverage. 
GAMA is funded by the STFC (UK), the ARC (Australia), the AAO, and the participating institutions. The GAMA website is \url{http://www.gama-survey.org/}.

To construct the GAMA-Mock the DiRAC Data Centric system at Durham University, operated by the Institute for Computational Cosmology on behalf of the STFC DiRAC HPC Facility (\url{www.dirac.ac.uk}), was used. 
This equipment was funded by BIS National E-infrastructure capital grant ST/K00042X/1, STFC capital grant ST/H008519/1, and STFC DiRAC Operations grant ST/K003267/1 and Durham University. DiRAC is part of the National E-Infrastructure. 
The development of the GAMA-Mock was supported by a European Research Council Starting grant (DEGAS-259586) and the Royal Society.

\emph{Software credit:}
We like to thank the developers and curators of the following software that this paper benefits from:
{\sc ipython} \citep{ipython}, {\sc matplotlib} \citep{matplotlib}, {\sc numpy} \citep{numpy},
{\sc pandas} \citep{pandas} and {\sc scipy} \citep{scipy}.


\bibliographystyle{mnras}
\bibliography{paper.bbl}

\begin{thebibliography}{}
\makeatletter
\relax
\def\mn@urlcharsother{\let\do\@makeother \do\$\do\&\do\#\do\^\do\_\do\%\do\~}
\def\mn@doi{\begingroup\mn@urlcharsother \@ifnextchar [ {\mn@doi@}
  {\mn@doi@[]}}
\def\mn@doi@[#1]#2{\def\@tempa{#1}\ifx\@tempa\@empty \href
  {http://dx.doi.org/#2} {doi:#2}\else \href {http://dx.doi.org/#2} {#1}\fi
  \endgroup}
\def\mn@eprint#1#2{\mn@eprint@#1:#2::\@nil}
\def\mn@eprint@arXiv#1{\href {http://arxiv.org/abs/#1} {{\tt arXiv:#1}}}
\def\mn@eprint@dblp#1{\href {http://dblp.uni-trier.de/rec/bibtex/#1.xml}
  {dblp:#1}}
\def\mn@eprint@#1:#2:#3:#4\@nil{\def\@tempa {#1}\def\@tempb {#2}\def\@tempc
  {#3}\ifx \@tempc \@empty \let \@tempc \@tempb \let \@tempb \@tempa \fi \ifx
  \@tempb \@empty \def\@tempb {arXiv}\fi \@ifundefined
  {mn@eprint@\@tempb}{\@tempb:\@tempc}{\expandafter \expandafter \csname
  mn@eprint@\@tempb\endcsname \expandafter{\@tempc}}}

\bibitem[\protect\citeauthoryear{{Aragon-Calvo}, {van de Weygaert}, {Jones}  \&
  {Mobasher}}{{Aragon-Calvo} et~al.}{2015}]{2015MNRAS.454..463A}
{Aragon-Calvo} M.~A.,  {van de Weygaert} R.,  {Jones} B.~J.~T.,   {Mobasher}
  B.,  2015, \mn@doi [\mnras] {10.1093/mnras/stv1903}, \href
  {http://adsabs.harvard.edu/abs/2015MNRAS.454..463A} {454, 463}

\bibitem[\protect\citeauthoryear{{Baldry} et~al.,}{{Baldry}
  et~al.}{2010}]{2010MNRAS.404...86B}
{Baldry} I.~K.,  et~al., 2010, \mn@doi [\mnras]
  {10.1111/j.1365-2966.2010.16282.x}, \href
  {http://adsabs.harvard.edu/abs/2010MNRAS.404...86B} {404, 86}

\bibitem[\protect\citeauthoryear{{Baldry} et~al.,}{{Baldry}
  et~al.}{2014}]{2014MNRAS.441.2440B}
{Baldry} I.~K.,  et~al., 2014, \mn@doi [\mnras] {10.1093/mnras/stu727}, \href
  {http://adsabs.harvard.edu/abs/2014MNRAS.441.2440B} {441, 2440}

\bibitem[\protect\citeauthoryear{{Baum}}{{Baum}}{1962}]{1962IAUS...15..390B}
{Baum} W.~A.,  1962, in {McVittie} G.~C.,  ed.,  IAU Symposium Vol. 15,
  Problems of Extra-Galactic Research. p.~390

\bibitem[\protect\citeauthoryear{{Beers}, {Flynn}  \& {Gebhardt}}{{Beers}
  et~al.}{1990}]{1990AJ....100...32B}
{Beers} T.~C.,  {Flynn} K.,   {Gebhardt} K.,  1990, \mn@doi [\aj]
  {10.1086/115487}, \href {http://adsabs.harvard.edu/abs/1990AJ....100...32B}
  {100, 32}

\bibitem[\protect\citeauthoryear{{Ben{\'{\i}}tez}}{{Ben{\'{\i}}tez}}{2000}]{2000ApJ...536..571B}
{Ben{\'{\i}}tez} N.,  2000, \mn@doi [\apj] {10.1086/308947}, \href
  {http://adsabs.harvard.edu/abs/2000ApJ...536..571B} {536, 571}

\bibitem[\protect\citeauthoryear{{Bilicki} et~al.,}{{Bilicki}
  et~al.}{2017}]{2017arXiv170904205B}
{Bilicki} M.,  et~al., 2017, preprint, \href
  {http://adsabs.harvard.edu/abs/2017arXiv170904205B} {} (\mn@eprint {arXiv}
  {1709.04205})

\bibitem[\protect\citeauthoryear{{Bolzonella}, {Miralles}  \&
  {Pell{\'o}}}{{Bolzonella} et~al.}{2000}]{2000A&A...363..476B}
{Bolzonella} M.,  {Miralles} J.-M.,   {Pell{\'o}} R.,  2000, \aap, \href
  {http://adsabs.harvard.edu/abs/2000A%26A...363..476B} {363, 476}

\bibitem[\protect\citeauthoryear{{Bonfield}, {Sun}, {Davey}, {Jarvis},
  {Abdalla}, {Banerji}  \& {Adams}}{{Bonfield}
  et~al.}{2010}]{2010MNRAS.405..987B}
{Bonfield} D.~G.,  {Sun} Y.,  {Davey} N.,  {Jarvis} M.~J.,  {Abdalla} F.~B.,
  {Banerji} M.,   {Adams} R.~G.,  2010, \mn@doi [\mnras]
  {10.1111/j.1365-2966.2010.16544.x}, \href
  {http://adsabs.harvard.edu/abs/2010MNRAS.405..987B} {405, 987}

\bibitem[\protect\citeauthoryear{{Boylan-Kolchin}, {Bullock}  \&
  {Kaplinghat}}{{Boylan-Kolchin} et~al.}{2011}]{2011MNRAS.415L..40B}
{Boylan-Kolchin} M.,  {Bullock} J.~S.,   {Kaplinghat} M.,  2011, \mn@doi
  [\mnras] {10.1111/j.1745-3933.2011.01074.x}, \href
  {http://adsabs.harvard.edu/abs/2011MNRAS.415L..40B} {415, L40}

\bibitem[\protect\citeauthoryear{{Boylan-Kolchin}, {Bullock}  \&
  {Kaplinghat}}{{Boylan-Kolchin} et~al.}{2012}]{2012MNRAS.422.1203B}
{Boylan-Kolchin} M.,  {Bullock} J.~S.,   {Kaplinghat} M.,  2012, \mn@doi
  [\mnras] {10.1111/j.1365-2966.2012.20695.x}, \href
  {http://adsabs.harvard.edu/abs/2012MNRAS.422.1203B} {422, 1203}

\bibitem[\protect\citeauthoryear{{Brammer}, {van Dokkum}  \& {Coppi}}{{Brammer}
  et~al.}{2008}]{2008ApJ...686.1503B}
{Brammer} G.~B.,  {van Dokkum} P.~G.,   {Coppi} P.,  2008, \mn@doi [\apj]
  {10.1086/591786}, \href {http://adsabs.harvard.edu/abs/2008ApJ...686.1503B}
  {686, 1503}

\bibitem[\protect\citeauthoryear{{Brown} et~al.,}{{Brown}
  et~al.}{2014}]{2014ApJS..212...18B}
{Brown} M.~J.~I.,  et~al., 2014, \mn@doi [\apjs] {10.1088/0067-0049/212/2/18},
  \href {http://adsabs.harvard.edu/abs/2014ApJS..212...18B} {212, 18}

\bibitem[\protect\citeauthoryear{{Brunner}, {Connolly}, {Szalay}  \&
  {Bershady}}{{Brunner} et~al.}{1997}]{1997ApJ...482L..21B}
{Brunner} R.~J.,  {Connolly} A.~J.,  {Szalay} A.~S.,   {Bershady} M.~A.,  1997,
  \mn@doi [\apjl] {10.1086/310674}, \href
  {http://adsabs.harvard.edu/abs/1997ApJ...482L..21B} {482, L21}

\bibitem[\protect\citeauthoryear{{Budav{\'a}ri}}{{Budav{\'a}ri}}{2009}]{2009ApJ...695..747B}
{Budav{\'a}ri} T.,  2009, \mn@doi [\apj] {10.1088/0004-637X/695/1/747}, \href
  {http://adsabs.harvard.edu/abs/2009ApJ...695..747B} {695, 747}

\bibitem[\protect\citeauthoryear{{Budav{\'a}ri}}{{Budav{\'a}ri}}{2012}]{2012amld.book..323B}
{Budav{\'a}ri} T.,  2012, {Photometric Redshifts: 50 Years After}.
??, pp 323--335

\bibitem[\protect\citeauthoryear{{Capak} et~al.,}{{Capak}
  et~al.}{2007}]{2007ApJS..172...99C}
{Capak} P.,  et~al., 2007, \mn@doi [\apjs] {10.1086/519081}, \href
  {http://adsabs.harvard.edu/abs/2007ApJS..172...99C} {172, 99}

\bibitem[\protect\citeauthoryear{{Carliles}, {Budav{\'a}ri}, {Heinis}, {Priebe}
   \& {Szalay}}{{Carliles} et~al.}{2010}]{2010ApJ...712..511C}
{Carliles} S.,  {Budav{\'a}ri} T.,  {Heinis} S.,  {Priebe} C.,   {Szalay}
  A.~S.,  2010, \mn@doi [\apj] {10.1088/0004-637X/712/1/511}, \href
  {http://adsabs.harvard.edu/abs/2010ApJ...712..511C} {712, 511}

\bibitem[\protect\citeauthoryear{{Cavuoti} et~al.,}{{Cavuoti}
  et~al.}{2017}]{2017MNRAS.466.2039C}
{Cavuoti} S.,  et~al., 2017, \mn@doi [\mnras] {10.1093/mnras/stw3208}, \href
  {http://adsabs.harvard.edu/abs/2017MNRAS.466.2039C} {466, 2039}

\bibitem[\protect\citeauthoryear{{Collister} \& {Lahav}}{{Collister} \&
  {Lahav}}{2004}]{2004PASP..116..345C}
{Collister} A.~A.,  {Lahav} O.,  2004, \mn@doi [\pasp] {10.1086/383254}, \href
  {http://adsabs.harvard.edu/abs/2004PASP..116..345C} {116, 345}

\bibitem[\protect\citeauthoryear{{Connolly}, {Csabai}, {Szalay}, {Koo}, {Kron}
  \& {Munn}}{{Connolly} et~al.}{1995}]{1995AJ....110.2655C}
{Connolly} A.~J.,  {Csabai} I.,  {Szalay} A.~S.,  {Koo} D.~C.,  {Kron} R.~G.,
  {Munn} J.~A.,  1995, \mn@doi [\aj] {10.1086/117720}, \href
  {http://adsabs.harvard.edu/abs/1995AJ....110.2655C} {110, 2655}

\bibitem[\protect\citeauthoryear{{Dahlen} et~al.,}{{Dahlen}
  et~al.}{2013}]{2013ApJ...775...93D}
{Dahlen} T.,  et~al., 2013, \mn@doi [\apj] {10.1088/0004-637X/775/2/93}, \href
  {http://adsabs.harvard.edu/abs/2013ApJ...775...93D} {775, 93}

\bibitem[\protect\citeauthoryear{{Driver} et~al.,}{{Driver}
  et~al.}{2011}]{2011MNRAS.413..971D}
{Driver} S.~P.,  et~al., 2011, \mn@doi [\mnras]
  {10.1111/j.1365-2966.2010.18188.x}, \href
  {http://adsabs.harvard.edu/abs/2011MNRAS.413..971D} {413, 971}

\bibitem[\protect\citeauthoryear{{Duarte} \& {Mamon}}{{Duarte} \&
  {Mamon}}{2015}]{2015MNRAS.453.3848D}
{Duarte} M.,  {Mamon} G.~A.,  2015, \mn@doi [\mnras] {10.1093/mnras/stv1799},
  \href {http://adsabs.harvard.edu/abs/2015MNRAS.453.3848D} {453, 3848}

\bibitem[\protect\citeauthoryear{{Duarte} \& {Mamon}}{{Duarte} \&
  {Mamon}}{2016}]{2016MNRAS.458.1301D}
{Duarte} M.,  {Mamon} G.~A.,  2016, \mn@doi [\mnras] {10.1093/mnras/stw389},
  \href {http://adsabs.harvard.edu/abs/2016MNRAS.458.1301D} {458, 1301}

\bibitem[\protect\citeauthoryear{{Duffy}, {Schaye}, {Kay}  \& {Dalla
  Vecchia}}{{Duffy} et~al.}{2008}]{2008MNRAS.390L..64D}
{Duffy} A.~R.,  {Schaye} J.,  {Kay} S.~T.,   {Dalla Vecchia} C.,  2008, \mn@doi
  [\mnras] {10.1111/j.1745-3933.2008.00537.x}, \href
  {http://adsabs.harvard.edu/abs/2008MNRAS.390L..64D} {390, L64}

\bibitem[\protect\citeauthoryear{{Firth}, {Lahav}  \& {Somerville}}{{Firth}
  et~al.}{2003}]{2003MNRAS.339.1195F}
{Firth} A.~E.,  {Lahav} O.,   {Somerville} R.~S.,  2003, \mn@doi [\mnras]
  {10.1046/j.1365-8711.2003.06271.x}, \href
  {http://adsabs.harvard.edu/abs/2003MNRAS.339.1195F} {339, 1195}

\bibitem[\protect\citeauthoryear{{Fontana}, {D'Odorico}, {Poli}, {Giallongo},
  {Arnouts}, {Cristiani}, {Moorwood}  \& {Saracco}}{{Fontana}
  et~al.}{2000}]{2000AJ....120.2206F}
{Fontana} A.,  {D'Odorico} S.,  {Poli} F.,  {Giallongo} E.,  {Arnouts} S.,
  {Cristiani} S.,  {Moorwood} A.,   {Saracco} P.,  2000, \mn@doi [\aj]
  {10.1086/316803}, \href {http://adsabs.harvard.edu/abs/2000AJ....120.2206F}
  {120, 2206}

\bibitem[\protect\citeauthoryear{{Furusawa}, {Shimasaku}, {Doi}  \&
  {Okamura}}{{Furusawa} et~al.}{2000}]{2000ApJ...534..624F}
{Furusawa} H.,  {Shimasaku} K.,  {Doi} M.,   {Okamura} S.,  2000, \mn@doi
  [\apj] {10.1086/308794}, \href
  {http://adsabs.harvard.edu/abs/2000ApJ...534..624F} {534, 624}

\bibitem[\protect\citeauthoryear{{Geha} et~al.,}{{Geha}
  et~al.}{2017}]{2017arXiv170506743G}
{Geha} M.,  et~al., 2017, preprint, \href
  {http://adsabs.harvard.edu/abs/2017arXiv170506743G} {} (\mn@eprint {arXiv}
  {1705.06743})

\bibitem[\protect\citeauthoryear{{Graham}, {Connolly}, {Ivezi{\'c}}, {Schmidt},
  {Jones}, {Juri{\'c}}, {Daniel}  \& {Yoachim}}{{Graham}
  et~al.}{2017}]{2017arXiv170609507G}
{Graham} M.~L.,  {Connolly} A.~J.,  {Ivezi{\'c}} {\v Z}.,  {Schmidt} S.~J.,
  {Jones} R.~L.,  {Juri{\'c}} M.,  {Daniel} S.~F.,   {Yoachim} P.,  2017,
  preprint, \href {http://adsabs.harvard.edu/abs/2017arXiv170609507G} {}
  (\mn@eprint {arXiv} {1706.09507})

\bibitem[\protect\citeauthoryear{{Guo} et~al.,}{{Guo}
  et~al.}{2011}]{2011MNRAS.413..101G}
{Guo} Q.,  et~al., 2011, \mn@doi [\mnras] {10.1111/j.1365-2966.2010.18114.x},
  \href {http://adsabs.harvard.edu/abs/2011MNRAS.413..101G} {413, 101}

\bibitem[\protect\citeauthoryear{{Hildebrandt} et~al.,}{{Hildebrandt}
  et~al.}{2012}]{2012MNRAS.421.2355H}
{Hildebrandt} H.,  et~al., 2012, \mn@doi [\mnras]
  {10.1111/j.1365-2966.2012.20468.x}, \href
  {http://adsabs.harvard.edu/abs/2012MNRAS.421.2355H} {421, 2355}

\bibitem[\protect\citeauthoryear{{Hildebrandt} et~al.,}{{Hildebrandt}
  et~al.}{2017}]{2017MNRAS.465.1454H}
{Hildebrandt} H.,  et~al., 2017, \mn@doi [\mnras] {10.1093/mnras/stw2805},
  \href {http://adsabs.harvard.edu/abs/2017MNRAS.465.1454H} {465, 1454}

\bibitem[\protect\citeauthoryear{{Hopkins} et~al.,}{{Hopkins}
  et~al.}{2013}]{2013MNRAS.430.2047H}
{Hopkins} A.~M.,  et~al., 2013, \mn@doi [\mnras] {10.1093/mnras/stt030}, \href
  {http://adsabs.harvard.edu/abs/2013MNRAS.430.2047H} {430, 2047}

\bibitem[\protect\citeauthoryear{Hunter}{Hunter}{2007}]{matplotlib}
Hunter J.~D.,  2007, Computing In Science \& Engineering, 9, 90

\bibitem[\protect\citeauthoryear{{Ilbert} et~al.,}{{Ilbert}
  et~al.}{2009}]{2009ApJ...690.1236I}
{Ilbert} O.,  et~al., 2009, \mn@doi [\apj] {10.1088/0004-637X/690/2/1236},
  \href {http://adsabs.harvard.edu/abs/2009ApJ...690.1236I} {690, 1236}

\bibitem[\protect\citeauthoryear{Jones, Oliphant, Peterson  et~al.}{Jones
  et~al.}{2001}]{scipy}
Jones E.,  Oliphant T.,  Peterson P.,   et~al., 2001, {SciPy}: Open source
  scientific tools for {Python}, \url {http://www.scipy.org/}

\bibitem[\protect\citeauthoryear{{Kafle}, {Sharma}, {Lewis}  \&
  {Bland-Hawthorn}}{{Kafle} et~al.}{2014}]{2014ApJ...794...59K}
{Kafle} P.~R.,  {Sharma} S.,  {Lewis} G.~F.,   {Bland-Hawthorn} J.,  2014,
  \mn@doi [\apj] {10.1088/0004-637X/794/1/59}, \href
  {http://adsabs.harvard.edu/abs/2014ApJ...794...59K} {794, 59}

\bibitem[\protect\citeauthoryear{{Kafle}, {Sharma}, {Lewis}, {Robotham}  \&
  {Driver}}{{Kafle} et~al.}{2018}]{2018MNRAS.475.4043K}
{Kafle} P.~R.,  {Sharma} S.,  {Lewis} G.~F.,  {Robotham} A.~S.~G.,   {Driver}
  S.~P.,  2018, \mn@doi [\mnras] {10.1093/mnras/sty082}, \href
  {http://adsabs.harvard.edu/abs/2018MNRAS.475.4043K} {475, 4043}

\bibitem[\protect\citeauthoryear{{Klypin}, {Kravtsov}, {Valenzuela}  \&
  {Prada}}{{Klypin} et~al.}{1999}]{1999ApJ...522...82K}
{Klypin} A.,  {Kravtsov} A.~V.,  {Valenzuela} O.,   {Prada} F.,  1999, \mn@doi
  [\apj] {10.1086/307643}, \href
  {http://adsabs.harvard.edu/abs/1999ApJ...522...82K} {522, 82}

\bibitem[\protect\citeauthoryear{{Kuijken} et~al.,}{{Kuijken}
  et~al.}{2015}]{2015MNRAS.454.3500K}
{Kuijken} K.,  et~al., 2015, \mn@doi [\mnras] {10.1093/mnras/stv2140}, \href
  {http://adsabs.harvard.edu/abs/2015MNRAS.454.3500K} {454, 3500}

\bibitem[\protect\citeauthoryear{{Laigle} et~al.,}{{Laigle}
  et~al.}{2016}]{2016ApJS..224...24L}
{Laigle} C.,  et~al., 2016, \mn@doi [\apjs] {10.3847/0067-0049/224/2/24}, \href
  {http://adsabs.harvard.edu/abs/2016ApJS..224...24L} {224, 24}

\bibitem[\protect\citeauthoryear{{Le Borgne} \& {Rocca-Volmerange}}{{Le Borgne}
  \& {Rocca-Volmerange}}{2002}]{2002A&A...386..446L}
{Le Borgne} D.,  {Rocca-Volmerange} B.,  2002, \mn@doi [\aap]
  {10.1051/0004-6361:20020259}, \href
  {http://adsabs.harvard.edu/abs/2002A%26A...386..446L} {386, 446}

\bibitem[\protect\citeauthoryear{{Leistedt} \& {Hogg}}{{Leistedt} \&
  {Hogg}}{2017}]{2017ApJ...838....5L}
{Leistedt} B.,  {Hogg} D.~W.,  2017, \mn@doi [\apj] {10.3847/1538-4357/aa6332},
  \href {http://adsabs.harvard.edu/abs/2017ApJ...838....5L} {838, 5}

\bibitem[\protect\citeauthoryear{{Libeskind}, {Guo}, {Tempel}  \&
  {Ibata}}{{Libeskind} et~al.}{2016}]{2016ApJ...830..121L}
{Libeskind} N.~I.,  {Guo} Q.,  {Tempel} E.,   {Ibata} R.,  2016, \mn@doi [\apj]
  {10.3847/0004-637X/830/2/121}, \href
  {http://adsabs.harvard.edu/abs/2016ApJ...830..121L} {830, 121}

\bibitem[\protect\citeauthoryear{{Liske} et~al.,}{{Liske}
  et~al.}{2015}]{2015MNRAS.452.2087L}
{Liske} J.,  et~al., 2015, \mn@doi [\mnras] {10.1093/mnras/stv1436}, \href
  {http://adsabs.harvard.edu/abs/2015MNRAS.452.2087L} {452, 2087}

\bibitem[\protect\citeauthoryear{{Loh} \& {Spillar}}{{Loh} \&
  {Spillar}}{1986}]{1986ApJ...303..154L}
{Loh} E.~D.,  {Spillar} E.~J.,  1986, \mn@doi [\apj] {10.1086/164062}, \href
  {http://adsabs.harvard.edu/abs/1986ApJ...303..154L} {303, 154}

\bibitem[\protect\citeauthoryear{{Mamon} \& {{\L}okas}}{{Mamon} \&
  {{\L}okas}}{2005}]{2005MNRAS.363..705M}
{Mamon} G.~A.,  {{\L}okas} E.~L.,  2005, \mn@doi [\mnras]
  {10.1111/j.1365-2966.2005.09400.x}, \href
  {http://adsabs.harvard.edu/abs/2005MNRAS.363..705M} {363, 705}

\bibitem[\protect\citeauthoryear{{Mamon}, {Biviano}  \& {Murante}}{{Mamon}
  et~al.}{2010}]{2010A&A...520A..30M}
{Mamon} G.~A.,  {Biviano} A.,   {Murante} G.,  2010, \mn@doi [\aap]
  {10.1051/0004-6361/200913948}, \href
  {http://adsabs.harvard.edu/abs/2010A%26A...520A..30M} {520, A30}

\bibitem[\protect\citeauthoryear{{Mamon}, {Biviano}  \& {Bou{\'e}}}{{Mamon}
  et~al.}{2013}]{2013MNRAS.429.3079M}
{Mamon} G.~A.,  {Biviano} A.,   {Bou{\'e}} G.,  2013, \mn@doi [\mnras]
  {10.1093/mnras/sts565}, \href
  {http://adsabs.harvard.edu/abs/2013MNRAS.429.3079M} {429, 3079}

\bibitem[\protect\citeauthoryear{{Matthews} \& {Newman}}{{Matthews} \&
  {Newman}}{2010}]{2010ApJ...721..456M}
{Matthews} D.~J.,  {Newman} J.~A.,  2010, \mn@doi [\apj]
  {10.1088/0004-637X/721/1/456}, \href
  {http://adsabs.harvard.edu/abs/2010ApJ...721..456M} {721, 456}

\bibitem[\protect\citeauthoryear{McKinney}{McKinney}{2012}]{pandas}
McKinney W.,  2012, Python for data analysis: Data wrangling with Pandas,
  NumPy, and IPython.
O'Reilly Media, Inc.

\bibitem[\protect\citeauthoryear{{Morrison}, {Hildebrandt}, {Schmidt},
  {Baldry}, {Bilicki}, {Choi}, {Erben}  \& {Schneider}}{{Morrison}
  et~al.}{2017}]{2017MNRAS.467.3576M}
{Morrison} C.~B.,  {Hildebrandt} H.,  {Schmidt} S.~J.,  {Baldry} I.~K.,
  {Bilicki} M.,  {Choi} A.,  {Erben} T.,   {Schneider} P.,  2017, \mn@doi
  [\mnras] {10.1093/mnras/stx342}, \href
  {http://adsabs.harvard.edu/abs/2017MNRAS.467.3576M} {467, 3576}

\bibitem[\protect\citeauthoryear{{Navarro}, {Frenk}  \& {White}}{{Navarro}
  et~al.}{1996}]{1996ApJ...462..563N}
{Navarro} J.~F.,  {Frenk} C.~S.,   {White} S.~D.~M.,  1996, \mn@doi [\apj]
  {10.1086/177173}, \href {http://adsabs.harvard.edu/abs/1996ApJ...462..563N}
  {462, 563}

\bibitem[\protect\citeauthoryear{{Navarro} et~al.,}{{Navarro}
  et~al.}{2004}]{2004MNRAS.349.1039N}
{Navarro} J.~F.,  et~al., 2004, \mn@doi [\mnras]
  {10.1111/j.1365-2966.2004.07586.x}, \href
  {http://adsabs.harvard.edu/abs/2004MNRAS.349.1039N} {349, 1039}

\bibitem[\protect\citeauthoryear{{Pawlowski}, {Ibata}  \&
  {Bullock}}{{Pawlowski} et~al.}{2017}]{2017arXiv171007639P}
{Pawlowski} M.~S.,  {Ibata} R.~A.,   {Bullock} J.~S.,  2017, preprint, \href
  {http://adsabs.harvard.edu/abs/2017arXiv171007639P} {} (\mn@eprint {arXiv}
  {1710.07639})

\bibitem[\protect\citeauthoryear{P\'erez \& Granger}{P\'erez \&
  Granger}{2007}]{ipython}
P\'erez F.,  Granger B.~E.,  2007, {C}omput. {S}ci. {E}ng., 9, 21

\bibitem[\protect\citeauthoryear{{Rahman}, {Mendez}, {M{\'e}nard}, {Scranton},
  {Schmidt}, {Morrison}  \& {Budav{\'a}ri}}{{Rahman}
  et~al.}{2016}]{2016MNRAS.460..163R}
{Rahman} M.,  {Mendez} A.~J.,  {M{\'e}nard} B.,  {Scranton} R.,  {Schmidt}
  S.~J.,  {Morrison} C.~B.,   {Budav{\'a}ri} T.,  2016, \mn@doi [\mnras]
  {10.1093/mnras/stw981}, \href
  {http://adsabs.harvard.edu/abs/2016MNRAS.460..163R} {460, 163}

\bibitem[\protect\citeauthoryear{{Robotham} et~al.,}{{Robotham}
  et~al.}{2010}]{2010PASA...27...76R}
{Robotham} A.,  et~al., 2010, \mn@doi [\pasa] {10.1071/AS09053}, \href
  {http://adsabs.harvard.edu/abs/2010PASA...27...76R} {27, 76}

\bibitem[\protect\citeauthoryear{{Robotham} et~al.,}{{Robotham}
  et~al.}{2011}]{2011MNRAS.416.2640R}
{Robotham} A.~S.~G.,  et~al., 2011, \mn@doi [\mnras]
  {10.1111/j.1365-2966.2011.19217.x}, \href
  {http://adsabs.harvard.edu/abs/2011MNRAS.416.2640R} {416, 2640}

\bibitem[\protect\citeauthoryear{{Robotham} et~al.,}{{Robotham}
  et~al.}{2012}]{2012MNRAS.424.1448R}
{Robotham} A.~S.~G.,  et~al., 2012, \mn@doi [\mnras]
  {10.1111/j.1365-2966.2012.21332.x}, \href
  {http://adsabs.harvard.edu/abs/2012MNRAS.424.1448R} {424, 1448}

\bibitem[\protect\citeauthoryear{{Sadeh}, {Abdalla}  \& {Lahav}}{{Sadeh}
  et~al.}{2016}]{2016PASP..128j4502S}
{Sadeh} I.,  {Abdalla} F.~B.,   {Lahav} O.,  2016, \mn@doi [\pasp]
  {10.1088/1538-3873/128/968/104502}, \href
  {http://adsabs.harvard.edu/abs/2016PASP..128j4502S} {128, 104502}

\bibitem[\protect\citeauthoryear{{Wojtak}, {{\L}okas}, {Gottl{\"o}ber}  \&
  {Mamon}}{{Wojtak} et~al.}{2005}]{2005MNRAS.361L...1W}
{Wojtak} R.,  {{\L}okas} E.~L.,  {Gottl{\"o}ber} S.,   {Mamon} G.~A.,  2005,
  \mn@doi [\mnras] {10.1111/j.1745-3933.2005.00054.x}, \href
  {http://adsabs.harvard.edu/abs/2005MNRAS.361L...1W} {361, L1}

\bibitem[\protect\citeauthoryear{{Wolf}}{{Wolf}}{2009}]{2009MNRAS.397..520W}
{Wolf} C.,  2009, \mn@doi [\mnras] {10.1111/j.1365-2966.2009.14953.x}, \href
  {http://adsabs.harvard.edu/abs/2009MNRAS.397..520W} {397, 520}

\bibitem[\protect\citeauthoryear{{Wright} et~al.,}{{Wright}
  et~al.}{2016}]{2016MNRAS.460..765W}
{Wright} A.~H.,  et~al., 2016, \mn@doi [\mnras] {10.1093/mnras/stw832}, \href
  {http://adsabs.harvard.edu/abs/2016MNRAS.460..765W} {460, 765}

\bibitem[\protect\citeauthoryear{{van der Walt}, {Colbert}  \&
  {Varoquaux}}{{van der Walt} et~al.}{2011}]{numpy}
{van der Walt} S.,  {Colbert} S.~C.,   {Varoquaux} G.,  2011, \mn@doi
  [Computing in Science \& Engineering] {10.1109/MCSE.2011.37}, 13

\makeatother
\end{thebibliography}

\bsp	
\label{lastpage}
\end{document}